\documentclass[12pt]{iopart}
\usepackage{bm}
\usepackage{graphicx}
\usepackage{xcolor}
\begin{document}

\title[Quantum imaging with sub-Poissonian light]{Quantum imaging with sub-Poissonian light: challenges and perspectives in optical metrology}
\author{I. Ruo Berchera, I. P. Degiovanni}
\address{INRIM, Strada delle Cacce 91, I-10135 Torino, Italy}
\ead{i.ruoberchera@inrim.it}

\begin{abstract}
Non-classical correlations in optical beams offer the unprecedented opportunity of surpassing conventional limits of sensitivity and resolution in optical measurements and imaging, especially but not only, when a low photon flux down to the single photons are measured. We review the principles of quantum imaging and sensing techniques which exploit sub-Poissonian photon statistics and non-classical photon number correlation, presenting some state-of-the-art achievements in the field. These quantum photonics protocols have the potential to trigger major steps in many applications, such as microscopy and biophotonics and represent an important opportunity for a new deal in radiometry and photometry.
\end{abstract}

\maketitle

\section{Introduction}

The technological exploitation of quantum states and quantum correlation, aiming to overcome the limits of conventional systems \cite{Dowling2003},  is one of the most, if not the most,  active research frontier nowadays. Scientists, from all the areas, are committed to apply these new paradigms to different physical platforms, from atomic  systems to solid state and photonic devices \cite{OBrien2009}. A tremendous scientific impact, ranging from biology and medicine to fundamental physics, and extraordinary technological advancement, ranging from communication and computation to precision sensing, are expected in few years or in a near future perspective. Quantum optics and quantum photonics are rather mature in this sense, including several technologies  already having a potential market diffusion, such as e.g. quantum key distribution \cite{Curty2014}.

In particular, quantum metrology, one of the main pillars of quantum technologies, has recently demonstrated to improve the sensitivity of one of the most sophisticated optical instruments currently available \cite{Aasi2013}, i.e. the large scale interferometers for gravitational wave detection, otherwise limited by the photon shot-noise.  Other examples of promising quantum enhanced measurement techniques, have been developed for particle tracking in optical tweezers \cite{Taylor-2014}, in sub-shot-noise wide field microscopy \cite{Samantaray2017}, quantum correlated imaging \cite{morris2015} and spectroscopy \cite{Kalashnikov2016}, displacement measurement \cite{Pooser2015} and remote detection and ranging \cite{Zhang-2015}.

In this process, the metrology community has a double role. On one side, the development of quantum technology needs a metrological infrastructure for the characterisation of the quantum photonics devices, and for their certification. This requires a knowledge of the basic principle of the quantum strategy, the expertise in the sources characterisation (e.g. their photon number statistics, the squeezing or sub-shot-noise properties, up to entanglement quantification), and of the detectors operating at the single- or  few-photon level. On the other side, the optical metrology has the opportunity to exploit the peculiar properties of quantum light to develop more accurate measurement, imaging and sensing techniques \cite{Genovese2016,review17}.

The process has already started: for example, heralded single photon source has been successfully applied to the calibration of single-photon detectors by metrologists \cite{Polyacov2007}, and similar quantum technique has been extended to low photon fluxes \cite{Brida2010OE, Avella2016}.

Of particular interest, from the radiometric point of view, is the possibility of developing absolute light sources with sub-shot-noise performances, since, the shot-noise level becomes a serious limitation to the uncertainty reduction in measurements performed in the few-photon regime. Probing and imaging delicate systems using small number of photons with true and significant sensitivity improvement would be extraordinarily important, for instance in biological and biochemical investigation \cite{Taylor2016}. Even more, there are biological effects which are triggered by a single photon, like the retina photo-transduction \cite{Phan2014}.  Thus, the quantum measurement strategy presented here, can be considered the precursors also of a potentially brand new metrological research field, namely quantum photometry.

In Sec. \ref{Limit of Classical (conventional) imaging}, we first present the fundamental limits of conventional (classical) optical measurements and imaging techniques, in particular the  Diffraction limit (DL) and the shot-noise  limit (SNL). Then, in Sec. \ref{Beyond classical limits by quantum states} we describe the properties of quantum states sources, such as the single photon source, Fock states and twin-beam, which, by virtue of their non-classical photon statistics and correlations allow surpassing classical limits. A certain number of quantum imaging techniques, their advantages and their limitations, are presented in Sec. \ref{Quantum imaging}. In Sec. \ref{Quantum Photometry}, we discuss single photon metrology for new directions in vision research and photometry. Short- and mid-term perspective in (quantum) optical metrology, and conclusions, are drawn in Sec. \ref{Final Remarks and Conclusion}.

\section{Limits of Classical (conventional) imaging}\label{Limit of Classical (conventional) imaging}

Measuring changes in intensity or in phase of an electromagnetic field, after interacting with matter, is the most simple way to extract relevant information on the properties of a system under investigation, whether a biological sample  \cite{biomedical0} or a digital memory disc \cite{Gu2014}.

The term "imaging" usually (not always) refers to the reconstruction of the whole spatial properties of the sample, namely in $2D$ or $3D$, and it can be achieved in two different modalities, wide field or point-by-point scanning. Wide field imaging is preferable in many cases, since it provides a more compete dynamic pictures but, for static sample, the point by point scanning offers advantages, for example the better $z$-resolution in confocal microscopy.

In any case, the two parameters quantifying the quality (the amount of information) of the image are the resolution, i.e. the minimum distance at which two points can be distinguished, and the sensitivity, i.e. the minimum measurable variation of the physical quantity in a certain point.

The quality of an image is always affected by several limitations, some of them avoidable by a careful design of the experiment (aberration, background, artifacts), others imposed by technical limitations of the available actual technology (for instance unavoidable noise or low efficiency of the detector), and others related to more fundamental reasons.
In particular diffraction limit and shot-noise limit represent "fundamental" bounds,  to resolution and sensitivity, at least when classical states of light are considered.
A possibility to overcome these limitations is offered by peculiar properties of quantum light.

\subsection{Diffraction Limit (DL)}

The diffraction limit, $R\simeq0.61 \lambda/NA$ (with $\lambda$ being the wavelength of the light and $NA$ the numerical aperture of the imaging system), represents the maximum obtainable imaging resolution in classical far-field imaging/microscopy. Depending on the kind of light radiation involved, namely incoherent or coherent, it takes the name of Abbe or Rayliegh diffraction limit, respectively.

The diffraction limit provides a lower bound to the current capability of precisely measuring the position of objects, from the small one such as e.g. single photon emitters (colour centres, quantum dots, etc.) \cite{Kurtsiefer00,Zaitsev00,Aharonovich09,Migdall13,Eisaman11}, to distant stars. In general,  the research of methods to obtain imaging and microscopy resolution below the diffraction limit is a topic of the utmost interest \cite{Perez12,Thiel07,Oppel12,Beskrovny05, D'Angelo01,Distante13,Bjork01} that could provide dramatic improvement in the observation of several systems spanning from quantum dots to living cells \cite{Taylor-2014,Igarashi12,LeSage12,Steinert13,Dolde13}, to distant astronomical objects. As a notable example, in several entanglement-related experiments using strongly coupled-single photon emitters it is of the utmost importance to measure their positions with the highest spatial resolution.
In principle, this limitation is overcome in microscopy by recently developed techniques such as e.g. Stimulated Emission Depletion (STED) and Ground State Depletion (GSD) \cite{Hell94,Hell95}, leading their inventors to win the Noble Prize in 2014. Nevertheless, even if they have been demonstrated effectively able to provide super-resolved imaging in a lot of specific applications, among which colour centres in diamond \cite{Rittweger09}, they are characterized by rather specific experimental requirements (dual laser excitation system, availability of luminescence quenching mechanisms by stimulated emission, non-trivial shaping of the quenching beam, high power).

In this review we will focus more on proposal and experiments of sub-diffraction imaging techniques exploiting the peculiar properties of quantum light, rather the ones connected to structured light and individually addressed or quenched fluorophores.

\subsection{Shot-Noise Limit (SNL)}\label{Shot-Noise Limit}

The sensitivity bound is established by the laws of quantum mechanics \cite{Giovannetti2011,Demkowicz2015,pirandola2017,paris2007} and is related to the mean energy of the probe beam.  In particular, in standard imaging and sensing exploiting classical probes, the sensitivity is fundamentally lower bounded by the "shot-noise limit" (SNL), $U_{SNL}\sim\langle N_{P} \rangle^{-1/2}$, where $\langle N_{P} \rangle$ is the mean number of photons in the probe.

Since the definition of classicality is not univocal, here we intend as "classical states" those providing experimental outcomes that would be completely explainable within the semiclassical theory of photodetection, in which the light is treated as a electromagnetic wave and the photo-current is discrete, being a flow of electrons \cite{Mandel95}. In the semiclassical picture, the probability of promoting an electron in the conduction band in a infinitesimal time interval is proportional to the light intensity. As a consequence, a plane wave with constant intensity  in macroscopic time interval generates $n_e$ photo-electrons  following  a Poissonian probability distribution $P(n_e)$.

Indeed, in quantum mechanics a plane wave with constant intensity is represented by a coherent state of the form:
\begin{equation}
|\beta\rangle = e^{|\beta|^2/2}\sum_{n=0}^\infty\frac{\beta^n}{\sqrt{n!}}|n\rangle,
\end{equation}
with Poissonian photon number statistics $P(n)= |\langle \hat{n}|\beta \rangle|^2= e^{-\langle\hat{n}\rangle^2}\frac{\langle \hat{n}\rangle^n}{n!}$, which reproduces the observed photocurrent by the absorption of a photon and the excitation of a photo-electron.
All the states that can be represented as a statistical mixture of coherent states, in  Glauber-Sudarshan  form
\begin{equation}\label{G-S}
\rho=\int d^2 \beta P(\beta)|\beta\rangle\langle\beta|, ~\mathrm{with }~ P(\beta)\geq0,
\end{equation}
are "classical" according to the definition conventionally used in quantum optics \cite{Davidovich1996}.
It is easy to show that these states must have Poissonian (or super-Poissonian) photon number fluctuation, with photon-number variance  $\langle\Delta^2\hat{n}\rangle$ equal (or larger) to the mean photon number, i.e. $\langle\Delta^2\hat{n}\rangle \geq\langle\hat{n}\rangle$.

\textit{By contrary, the detection of sub-Poissonian variance, $\langle\Delta^2\hat{n}\rangle < \langle\hat{n}\rangle$, is sufficient to indicate the  non-classical nature of the light, according to the quantum optics definition.}

\begin{figure}[htbp]
	\centering
	\includegraphics[ width=0.8\textwidth, angle=-90 ]{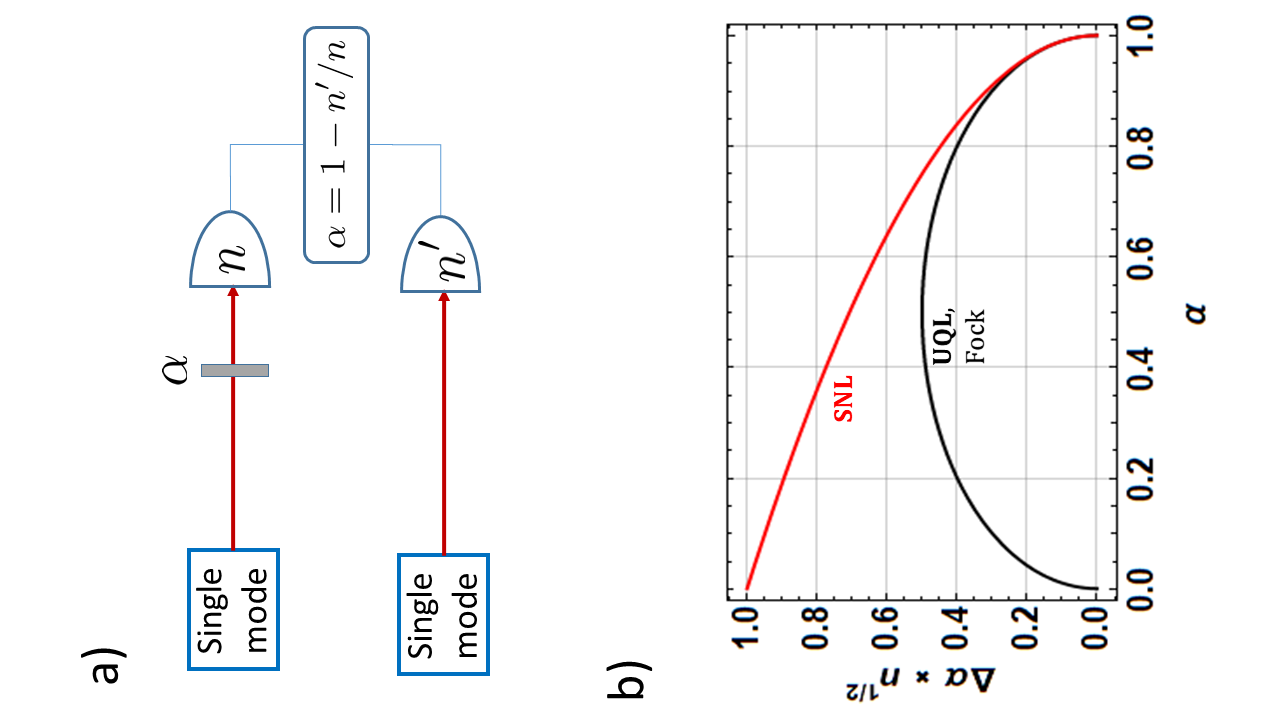}
	\caption{Loss estimation. (a) The loss estimation is obtained by comparing the number of photons detected with and without the absorbing sample. (b) Uncertainty in function of the loss parameter $\alpha$. SNL stands for shot-noise limit, while UQL stands for ultimate quantum limit}\label{D_alpha_Fig}
\end{figure}

In order to study the photon statistics influence on the optical measurement, let us consider the problem of absorption estimation. Sending a probe beam with mean number of photon $\langle\hat{n}\rangle$ through a sample with absorption factor  $\alpha$, and measuring the beam mean power after the interaction one get $\langle\hat{n'}\rangle=(1-\alpha) \langle\hat{n}\rangle$, see Fig. \ref{D_alpha_Fig}(a). By uncertainty on the absorption is simply given by:
 \begin{equation}\label{unc_prop}
 \Delta \alpha=\frac{\sqrt{\langle\Delta^{2}  \hat n' \rangle}}{\left|\frac{\partial \langle n'\rangle}{\partial\alpha}\right|}.
 \end{equation}
If the probe statistics is Poissonian  immediately one obtains  $\Delta \alpha_{SNL}=[(1-\alpha)/ \langle\hat{n}\rangle]^{1/2}$, graphically represented by the red line in Fig. \ref{D_alpha_Fig}(b). Thus, it is clear how the SNL is related to the Poissonian fluctuation in classical states of light.

Phase estimation problem can be treated similarly. The output power of a Mach Zehnder or a Michelson interferometer is $\langle\hat{n'}\rangle=\tau \langle\hat{n}\rangle$, where $\tau$ depends on the phase shift $\phi$ among the arms, $\tau=\sin^{2}(\phi/2)$.  Using the uncertainty propagation on Eq. (\ref{unc_prop}) where $\phi$ is placed instead of $\alpha$, and assuming Poissonian statistics of the light, one gets  $\Delta \phi= \langle\hat{n}\rangle^{-1/2} \cos(\phi/2)^{-1}$. For $\phi=0$ it reaches the lowest value, which is again the SNL.

Such classical limits in loss and phase estimations that we have found here by assuming a specific and simple detection strategy and Poissonian light, coincide with the ultimate bounds achievable by a coherent state and the most general measurement allowed by quantum mechanics: see \cite{paris2007,adesso2009,illuminati} for loss estimation and \cite{Giovannetti2011,Demkowicz2015,Giovannetti04} for phase estimation.

Of course, increasing arbitrary the number of photons, i.e. the optical power, one can always reduce the shot-noise contribution to the uncertainty budget below other technical noise. However, this is not always an option.
Beating such a limit is particularly important when there are optical power constraints. In addition to fundamental problems related to  quantum back-action \cite{Giovannetti04}, there are several practical situations in which one may need to limit the optical power:

\begin{itemize}
	
\item 	\textit{Extremely precise measurements:} In the last generation of gravitational wave detector the power circulating in the interferometer is of the order of 1 kW (CW), about $10^{21}$ Photon/s. This allows to reach the sensitivity in the strain of $\Delta h<10^{-22}$. A further increasing of the power would have thermal effects on the surface of the mirrors  and other optical components, producing an unwanted large scattering interfering with beam propagation.
\item \textit{ Probing delicate systems (biological sample, photosensitive chemicals)}: Damages due to optical tweezers have been reported for Escherichia coli, Listeria and other bacteria. Alteration of chemical and biological photo sensitive processes can occur at e very low power. %\textbf{[REF!! necessaria? non c'e' negli altri item]}
\item \textit{Investigate response of a system at few photons level:} Some biophysical and biochemical processes, for example phototransduction in vision, or photosynthesis  are triggered by the absorption of one or few photons. Moreover, the development of quantum technologies are strongly based on single or few photon state manipulation and detection. In order to study and calibrate a single-photon detector, or the retina response at the few photons level, it is fundamental to develop sub-shot-noise techniques.

\item  \textit{Developing standards and traceable measurements at few photons/single photon level:}  the need for dedicated metrology tools in the few photons regime is one of the key for the success of the quantum photonics technologies (quantum cryptography, quantum sensing and imaging). On the other side, new insights in the study of the retinal process and response at low light level are a strong motivation for the development of quantum photometry.

\end{itemize}

\section{Beyond classical limits using quantum states}\label{Beyond classical limits by quantum states}

One can note that the conventional statistical scaling of the uncertainty after $N$ independent repetitions of the same measurement coincides with  the scaling of the SNL when $N$ photons are detected, $N^{-1/2}$. This means that photons in photons counting measurements behave somehow independently each other, which results in Poissonian behaviour.
However, quantum mechanics does not prevent a light beams to have sub-Poissonian fluctuations, or more in general with a strong degree of cooperation among photons, where the probability of detecting a photon at a certain time $t$ is correlated or anti-correlated to the detection of a photon at time $t_0$. Ordinate and pseudo-deterministic stream of photons, like produced by single-photon sources, and pairs of correlated beams named "twin beam" (TWB) can now be ordinarily generated.

For sake of completeness, aside sub-Poissonin statistics, there are other way to define the "quantumness" of a state, the most important of them is the concept of entanglement. The state of a system composed by two (or more) subsystems, for example a pair particles, is entangled \emph{if and only if} it cannot be written in terms of the product state of each individual subsystems. The entanglement implies the existence of a degree of correlation among the particles which cannot be explained in any realistic and local theory \cite{Genovese2005}. It has been definitely verified  that in an entangled state, the measurement of a particle influence the results of the measurement on the other particle even if they are space-like separated, i.e. causally disconnected \cite{Giustina2015,Shalma2015}, revealing the existence of that kind of "spooky action at distance" in the words of Einstein \cite{epr}.

Most theoretical investigations in quantum metrology, have been addressed to improve the scaling of the uncertainty with the photon number exploiting entanglement, up to the ultimate limit imposed by quantum mechanics, $N^{-1}$, known as Heisenberg limit. For large photon number, Heisenberg scaling  would bring enormous advantage and many schemes have been proposed \cite{huelga, boto,ho,ca}, typically using the entangled  state of the form $2^{-1/2}(|n_A 0_B\rangle+ |0_A n_B\rangle)$ (NOON state). This state  is a quantum coherent superposition of two possibilities: either the $n$ photons pass all in the arm $A$ of an interferometer or they travel all in the arm $B$. A phase shift $\varphi$ acting of the arm $A$ introduces a global phase difference $n\varphi$ among the two possibilities of the superposition. Combining at a BS the two modes $A$ and $B$ gives a $n$-time denser interference fringes, $\sin^{2}(n\varphi)$, reaching the Heisenberg scaling of the sensitivity.
Unfortunately,  while two photon entangled states can be routinely produced by photon pairs emitted in Spontaneous Parametric Down Conversion (SPDC) and Hong Ou-Mandel effect (described in Sec. \ref{Fock states and single photon sources}), generation and detection of entangled states with a larger number of photons, i.e. $n>4$, is extremely challenging.

Even worse, entanglement itself is extremely fragile to the losses, for example loosing a single photon from  a NOON state projects it in a classical mixture. A real advantage is preserved only if $\eta^{n} v^{2}>1$, where $\eta$ is the detection efficiency and $v$ the visibility of the interference \cite{Slussarenko17}; this condition becomes harder and harder to fulfill at the increasing the photon number $n$.  Other quantum states, entangled and squeezed, which are more resilient to experimental imperfections \cite{Ja}, have been considered, but nevertheless reaching the Heisenberg limit for a large number of photons, is probably a chimera. In fact, recently it has been shown that in presence of decoherence  the Heisenberg limit, and in general any chance of the uncertainty scaling with the photon number, is out of reach \cite{Demkowicz-2012,i7}. Rather, the enhancement with respect to the standard quantum limit is given  by a constant factor, for example it takes the form  $\sqrt{(1-\eta)/\eta}$ in presence of a loss factor $(1-\eta)$ \cite{Demkowicz-2015,Tsang-2013}.

Nevertheless, the possibility of generating entangled states  (such as NOON states with N=2) and the availability of single-photon detectors have enabled the demonstration of the quantum states potentiality in super-resolved lithography \cite{D'Angelo01,Dema}, phase contrast polarization microscopy \cite{On,Is}, magnetic field sensing \cite{Wo} and solution concentration measurement \cite{Cr} .

On the other side, quantum advantages can be obtained more easily by exploiting non-classical Gaussian states \cite{i6}, which are relatively easy to be produced experimentally, such as squeezed vacuum generated by SPDC and Optical Parametric Oscillators (OPO). Single-mode squeezing \cite{Andersen-2016,Schnabel-2016} in one of the quadratures (generated by OPO) has been the first quantum property considered in quantum metrology, in particular for quantum enhanced interferometry \cite{Caves-1981}  and the more successful from the practical point of view, leading to a real sensitivity improvement of the modern gravitational wave detectors \cite{Schnabel-2016, Abbie-2011}. It leads also to promising application in the photonic force microscopy for biological particle tracking \cite{Taylor-2014,Taylor-2013} and beam displacement measurement \cite{Pooser2015,Treps2003}.

In the context of this review it is useful to dwell on how the optical losses modify the light properties. In quantum optics loss mechanisms are usually described by the action of a beam splitter (BS) with transmittance  $\tau$ and reflectance $1-\tau$, having the field of interest entering one of the input ports of the BS and the "vacuum" entering the other port. The linear unitary evolution leads to the following statistics of the transmitted photon number $n'$, in function of the input one $n$:

\begin{eqnarray}\label{phcStat}
\langle\hat{n'}\rangle = \tau \langle\hat{n}\rangle\\
\langle\Delta^2\hat{n'}\rangle=\tau^2 \langle\Delta^2\hat{n}\rangle+\tau(1-\tau)\langle\hat{n}\rangle \nonumber
\end{eqnarray}

Note that, in presence of losses, photon statistics always contain a shot-noise contribution, the one proportional to the mean value, regardless the input variance. This can be seen as the effect of the vacuum fluctuation at the unused port of the BS. From Eq.s ($\ref{phcStat}$), it is clear that the Poissonian fluctuation in input, $\langle\Delta^2\hat{n}\rangle=\langle\hat{n}\rangle$, remains Poissonian just by a rescaling of the man photon number. The thermal light (black-body),  which have super-Poissonian statistics such as $\langle\Delta^2\hat{n}\rangle=\langle\hat{n}\rangle(1+\langle\hat{n}\rangle)$,  remains unchanged too.

On the contrary, losses negatively affect the sub-Poissonian character of light, e.g. Fock states $|n\rangle$, eigenstates of the photon number operator, having (by definition) zero-fluctuation $\langle\Delta^2\hat{n}\rangle = 0$, have as output statistic in presence of losses  $\langle\Delta^2\hat{n'}\rangle=\tau(1-\tau)\langle\hat{n}\rangle$, which approaches the Poissonian one for $\tau\ll1$.

In Sec. \ref{Shot-Noise Limit} we have derived the uncertainty in loss estimation in case of Poissonian light. Now, substituting the quantities of Eq.s~(\ref{phcStat}) in the uncertainty expression in Eq. (\ref{unc_prop})  (with $\tau=1-\alpha$) we can derive the more general expression:
\begin{equation}\label{uncdr}
\Delta \alpha = \sqrt{\frac{\alpha(1-\alpha)+F (1-\alpha)^{2}}{\langle \hat{n}\rangle}},
\end{equation}
where $F=\langle\Delta^2\hat{n}\rangle/\langle\hat{n}\rangle$ is the Fano factor as it would be measured in absence of the object. For Poissonian probe, $F=1$, one retrieves the SNL, $\Delta \alpha_{SNL}$. However, sub-Poissonian light with $F<1$ allows surpassing the SNL. The most favorable case is when $\langle\Delta^2\hat{n}\rangle=0$  (Fock state),  leading to $\Delta \alpha_{Fock} = \sqrt{ \alpha} \Delta \alpha_{SNL}$  depicted in  Fig. \ref{D_alpha_Fig}(b), black line. Incidentally, it has been shown that the last one represents the ultimate quantum limit (UQL) in the loss estimation for a single mode interrogation of the sample \cite{paris2007,adesso2009} as well as for entangled bipartite states \cite{illuminati}. The advantage is dramatic for small absorption, a region which is particularly significant in many real applications, for example when imaging thin biological samples or detecting low density gas flowing or low concentration in solution.

\subsection{Twin-beam}\label{Twin-beam}

In practice, realizing a single mode beam with significant sub-shot-noise fluctuation reduction is quite challenging. Another way, more effective and commonly used in experimental demonstrations, relays upon the non-classical photon number correlation among two beams, the "twin-beam" (TWB) state. The idea is that one beam of the pair is used as a probe while the other acts as a reference for the quantum noise, a property that can be exploited from interferometry \cite{ruo5, ruo6} to imaging \cite{Genovese2016, review17, Kolobov-2007}.

\begin{figure}[thbp]
	\center
	\includegraphics[width=5in]{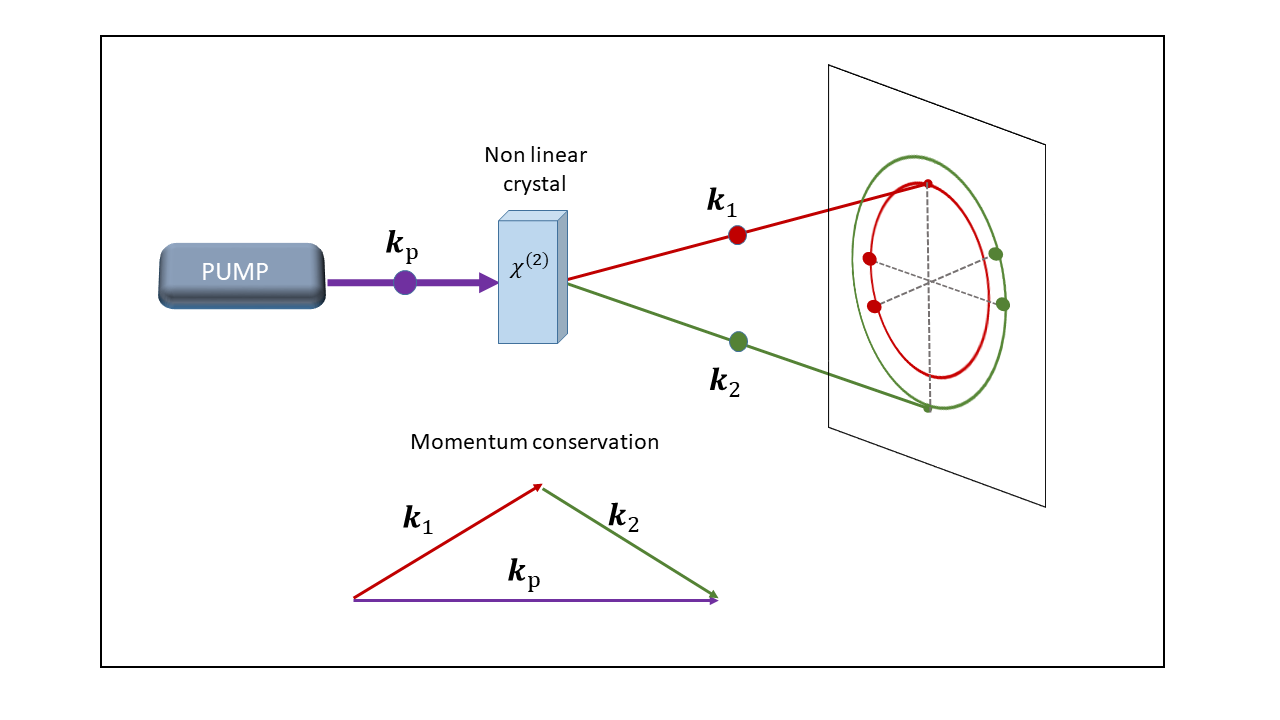}
	\caption{Generation of TWB state by SPDC: the process occurs inside a crystal with second order susceptibility. A photon of an input pump field is down-converted in a pair of lower energy photons conserving also the momenta (phase matching). The phase matching condition establishes a relation between the emission angle and the wavelength (frequencies) of the photons. Thus, correlated photons appear in opposite direction, along circles corresponding to their specific wavelength, according to the energy conservation. }\label{SPDC_Fig}
\end{figure}

SPDC is one of the most efficient ways to produce quantum correlations between optical fields. This physical phenomenon was discovered at the end of the sixties \cite{ZEL69,BUR70} and occurs due to the interaction between an intense optical field, usually called pump beam, and a non-linear dielectric optical medium \cite{Aspdc2}.  Basically, the phenomenon consists in the decay of one photon of the pump into two photons, named "signal" and "idler" for historical reasons, preserving energy and momentum (see Fig. \ref{SPDC_Fig}):

\begin{eqnarray}
\omega_p &=&  \omega_1 + \omega_2 \nonumber \\
\textbf{k}_p &=&  \textbf{k}_1 + \textbf{k}_2
\label{phasematch}
\end{eqnarray}

where $\omega_p$ is the frequency of the pump photon and $\omega_1$, $\omega_2$ are the frequencies of the signal and idler photons, and $\textbf{k}_j$ (with $j=p,1,2$) are the  corresponding wave vectors.
Even though the emission of a pair is a quantum random process, not different from photon emission by a thermal source, the presence of one photon with a certain direction and frequency is bound to the presence of a "twin" photon in a correlated spatial-frequency mode. In the high gain regime of SPDC, many photon pairs can be also generated occupying plethora of bipartite correlated modes \cite{review17, Aspdc2}. Approximatively this corresponds to the parallel generation of many entangled states, each formally written as:

\begin{equation}
\vert TWB\rangle_{1,2}=\sum_{n} c_{n}\vert n\rangle_{1}\vert n\rangle_{2},
\label{SPDC_state}
\end{equation}

where for simplicity the subscript "1" represents the spatial-frequency mode $(\textbf{q},\omega)$ and the subscript "2" represents the correlated mode $(-\textbf{q},\omega_{p}- \omega)$ ($\textbf{q}$ is the transverse momentum) \cite{Aspdc2}. The probability amplitude is $c_{n } \propto \sqrt{\mu^n/(\mu + 1)^{n+1}}$, where $\mu$ is the mean number of photon per mode.

From the form of the coefficients $c_{n}$ it emerges the super-Poissonian (thermal) character of the pairs emission. However, the thermal fluctuations are perfectly reproduced in the two modes. The degree of correlation is quantified by the noise reduction factor $\sigma$, as the ratio between the variance of the difference in the number of photons in the two modes, normalized to the corresponding shot-noise  \cite{Brida2010OE, Je04, Mo05, Bl08, Bo07, Pe12, Is16, Is16b, Is09, Ta91, La14}:
\begin{equation}\label{NRF}
\sigma=\frac{\langle\Delta^2(\hat{n_1}-\hat{n_2})\rangle}{\langle\hat{n_1}+\hat{n_2}\rangle}\equiv\frac{\langle\Delta^2\hat{n_1}\rangle+\langle\Delta^2\hat{n_2}\rangle-2 \langle\Delta\hat{n}_1\Delta\hat{n}_2\rangle}{\langle\hat{n_1}+\hat{n_2}\rangle}
\end{equation}
For classical states, $\sigma$ is lower bounded by 1, while for TWB in the form of Eq. (\ref{SPDC_state}) it is $ \sigma_{TWB} =0$. Also in this case the practical limit in the noise reduction is usually represented by the unavoidable presence of losses. From Eq. (\ref{phcStat}), considering two modes subject to the same transmission-detection efficiency $\eta_1=\eta_2=\eta$,
\begin{equation}\label{NRF_eta}
\sigma_{det} = \eta \sigma+ 1-\eta. \label{13}
\end{equation}
The lower bound in presence of losses is therefore $\sigma_{det} =  1-\eta$.

Sub-shot-noise (SSN) correlation of this state has been experimentally demonstrated both in the case of a single twin beam as in Eq. (\ref{SPDC_state}) \cite{Bo07,Is16,Heidmann1987,Mertz1990,agafonov2011} and in the case of many spatial modes detected in parallel by the pixels of a CCD camera \cite{Je04,Bl08,brida2009}.

\begin{figure}[thbp]
	\center
	\includegraphics[width=3.5in, height=2.5in, angle=-90]{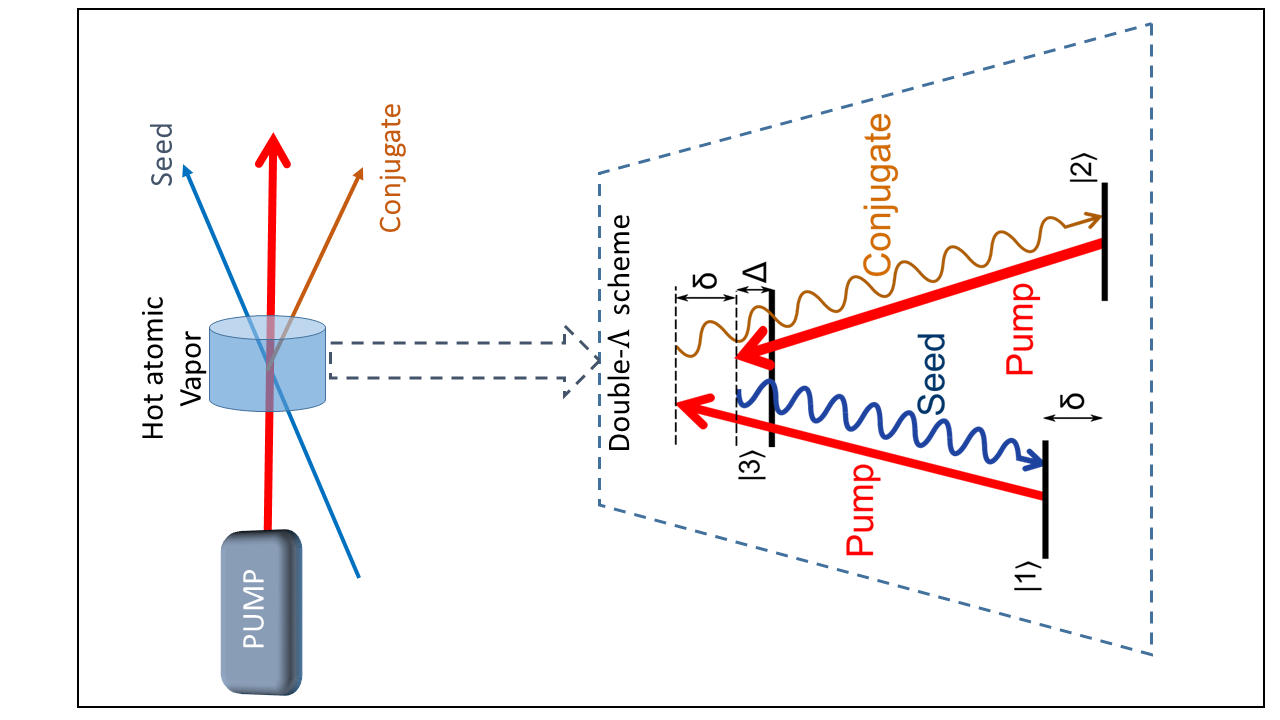}
	\caption{Generation of TWB stateby FWM: In this process two degenerate photons of a pump beam are absorbed in a hot atomic vapor cell operated in a double-$\Lambda$ configuration,  promoting the emission of two new photons with different energies. The emission is usually stimulated by injecting a seed beam, which is amplified together with the emission of a macroscopic conjugate beam.}\label{FWM_Fig}
\end{figure}

SSN correlations can be efficiently generated also by four-wave-mixing (FWM) in atomic vapours \cite{Glorieux2011,Embrey2015,cao2017, boyer2008}, a process depicted in Fig.\,\ref{FWM_Fig}. A double-$\Lambda$ configuration  with three atomic levels is exploited (often the hyperfine splitting structure of Rubidium), so that when two pump photons are absorbed, two photons, the signal and  the conjugate, are generated ($\omega_{conj}=\omega_{sign}+2\delta$, where $\delta$ is the hyperfine energy splitting). Given the intrinsic high gain of the process, and the stimulation obtained by seeding the signal with a macroscopic beam, FWM leads to the production of intense quantum correlated beams that can reach noise reduction $\sigma <1$ for several tens of $\mu$W of optical power.

Based on this strong non classical correlation, TWB states have shown the possibility of sub-SNL sensitivity in absorption/transmission measurements \cite{Ta91,rarity,Hayat1999, Moreau2017,Losero2018}, quantum ellipsometry \cite{Abouraddy2001,Toussaint2004},  quantum enhanced sensing \cite{Pooser2015,Zhang-2015,lopaeva2013, clark2012}, quantum reading of digital memories \cite{pirandola2011} and plasmonic sensors \cite{lawrie2013, pooser2016}.
The common idea behind these works is that the random intensity noise in the probe beam addressed to the sample can be known by measuring the correlated (reference) beam and subtracted.

%Note that the two-beams approach is extensively used in standard devices like spectrophotometers, where a classical beam is split in two by a beam splitter and one beam is used to monitor the instability of the source and detectors and to compensate for them.  This is particularly effective in practical applications, since unavoidable drifts in the source emission or detector response would lead to strong bias, especially in the estimation of small absorptions. However, in classical correlated beams (CCB) generated in this way, only the super-Poissonian component of the fluctuations is correlated (sometimes called classical "excess noise"),  whereas the shot noise remains uncorrelated and cannot be compensated. Therefore TWB represent the natural extension to the two-beam approach to the quantum domain, promising to be especially effective for small absorption measurement and when low photon flux is required.

These considerations can be extended to the multi-mode case. Indeed, when TWB are produced through %traveling wave
 parametric down conversion, or by FWM in atomic vapors \cite{boyer2008,Corzo-2012, adenier}, the emission is spatially broadband, a collection of pairwise correlated modes in the transverse plane. Modern high sensitivity multi-pixel detectors, like Charge Coupled Device (CCD), Complementary Metal-Oxide Semiconductor (CMOS) cameras, or single-photon detector array, can detect simultaneously thousands of correlated spatial modes with high efficiency, improving the sensitivity of imaging applications even in wide-field modality \cite{Samantaray2017,gatti2008,Brida2010,brida2011-1}.

One of first application of SPDC entangled photons has been the "ghost imaging" (GI)\cite{belinskii,pittman}, whose goal is the reconstruction of the spatial transmission/reflection profile of an object even if the interacting photons are collected by a single pixel detector. After the first realization,it has been shown that GI can be obtained  also with classical beams, nevertheless,  non-classical correlations provides some advantage at very low photon flux \cite{Gatti2004, Erkmen2009, brida2011-2}.

Finally, non-classical correlations have disclosed new possibilities in quantum radiometry \cite{Zwinkels2010}, e.g. the possibility of absolute calibration of detectors without the need of comparison with calibrated standards. The first proposal for calibrating single-photon detectors has been formulated by Klyshko \cite{ZEL69} just after the discover of SPDC process and nowadays it is an established technique \cite{Polyacov2007}, currently used in metrological institutes. Generalizing the method to the domain of analog detectors and spatially resolving detector has recently lead to the  absolute calibration of electron multiplying CCD cameras (EMCCD) from to the linear regime \cite{Brida2010OE,Meda2014} to the on-off single photon regime \cite{Avella2016}, just changing the intensity of the SPDC pump laser, as well as of ICCD cameras \cite{iccd, iccd2}.

\subsection{Fock states and single photon sources}\label{Fock states and single photon sources}
Single-photon sources and more in general Fock state sources are necessary for applications  in the field of quantum technologies
(e.g. quantum computing, quantum key distribution and quantum metrology), which are among the most relevant topics with respect to innovation and high technology worldwide.
An ideal single-photon source emits one photon on demand, at a time chosen by the user, with the emitted photons being indistinguishable from one another and having an adjustable repetition rate \cite{Lounis05,Scheel09, Polyakov09}.

\begin{figure}[tbp]
	\centering
	\includegraphics[ width=0.8\textwidth ]{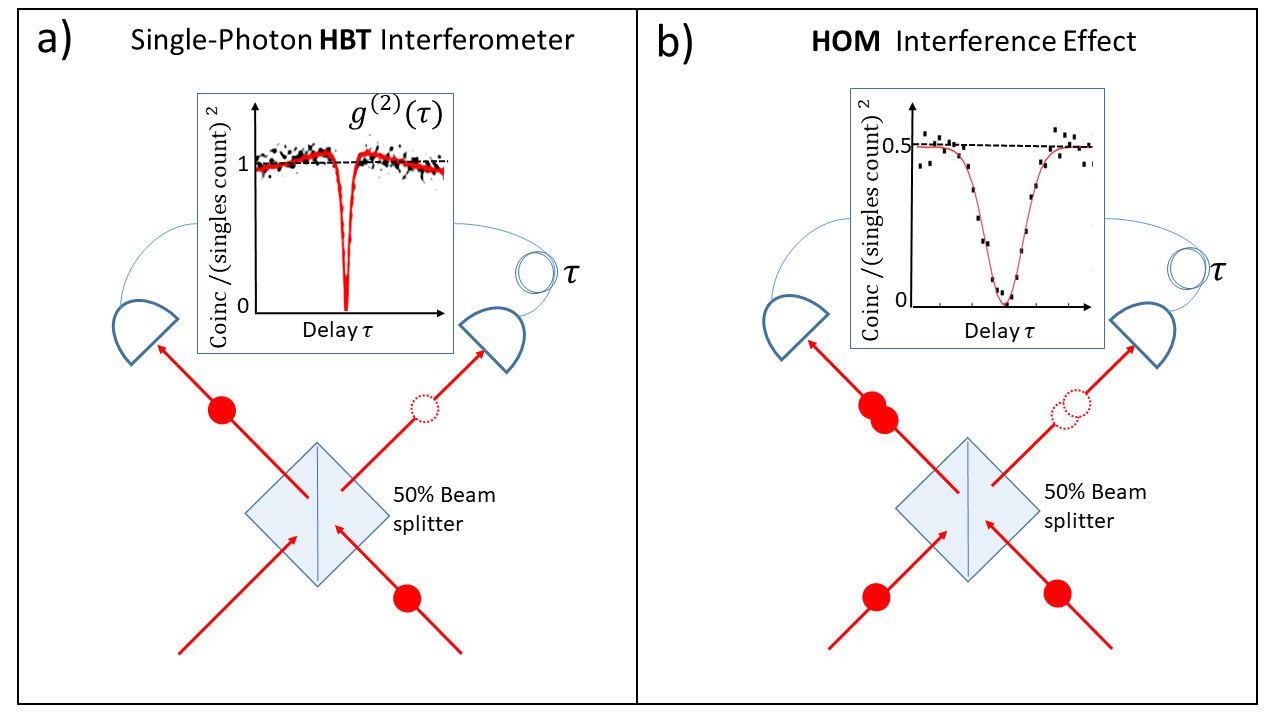}
	\caption{Figures of merit for characterizing single-photon sources. a) Sketch  of the Hanbury-Brown and Twiss experiment. A photon impinges a 50\% beam splitter: in half of the cases the photon is reflected ($p_{R}=1/2$) and in the other half it is transmitted ($p_{T}=1/2$), no chance to have a coincidences among two detectors. If a subsequent photon arrives after a time interval $\tau\neq0$ there is a probability $1/2= p_{R}\cdot p_{T}+p_{T}\cdot p_{R}$
% $1/2= p_{R}\; p_{T}+p_{R}\; p_{T}$
that the two photons are registered by different detectors. b) Hong-Ou-Mandel effect. If two indistinguishable photons arrives at the same time ($\tau=0$) at the ports of the beam splitter, a quantum interference phenomenon occurs due to the fact that the amplitude probabilities $\mathcal{A}_{TT}$ and $\mathcal{A}_{RR}$ of the events of the type $TT$ (both the photon transmitted) and $RR$ (both the photon reflected)  respectively, sum up coherently, i.e. with their quantum phases. Since each reflected photon acquire a phase $\pi/2$, the amplitude probability $\mathcal{A}_{RR}$  is equal but opposite in sign to $\mathcal{A}_{TT}$ (because $\exp{\pi/2}*\exp{\pi/2}=-1$). Thus, the coincidence probability $|\mathcal{A}_{TT}+\mathcal{A}_{RR}|^{2}$ is null. }\label{HBT_HOM_Fig}
\end{figure}

Based on these requirements, two figures of merit are generally used for characterizing the quality of a single photon source. They are represented in Fig. \ref{HBT_HOM_Fig}. The Hanbury-Brown and Twiss (HBT) interferometer in Fig. \ref{HBT_HOM_Fig}(a) measures the tendency of the photons to arrive in pairs.  Formally, the HBT experiment allows the evaluation of the second order Glauber correlation function, defined as $g^{(2)}(\tau=0)\equiv\langle \hat n(\hat n-1)\rangle/\langle \hat n\rangle^{2}$. For coherent light (a stable laser) this parameter equals the unity, for thermal light it turns out to be 2, while it must be zero if the photons are emitted one by one. Thus, for a single-photon source, the coincidences level among the two detectors at the varying of the temporal delay $\tau$ presents a dip exactly at $\tau=0$. The second scheme, sketched in Fig.\ref{HBT_HOM_Fig}(b) and known as Hong-Ou-Mandel (HOM) interferometer, measures the indistinguishability of two photons produced at different times by a single photon source. Only if the photons (or more precisely their paths) cannot be distinguished, even in principle, with respect to any degrees of freedom (polarization, wavelength, spatial mode ect..), the quantum interference happens, forcing the photons to exit always from the same side of the beam splitter and no coincidences are registered. If the two photons can be identified, for example temporally, the interference disappear. Therefore, only for $\tau=0$ a dip is observed.

A Fock state source \cite{Mandel95} is a generalisation of a single photon source, since it should be able to emit a fixed number of photons on demand and  indistinguishable from one another.
Obviously, a Fock state source emitting $n$ photons per pulse can be realised exploiting $n$ ideal single-photon sources operating together.
Such photon sources have the potentiality to become a new quantum standard with a huge range of applications: calibration/characterisation of
single photon counter devices \cite{Lundeen09, Avella11, Brida12, Brida12b}, realization of the SI base unit candela \cite{Zwinkels2010}, quantum enhanced measurements \cite{Giovannetti04}, quantum sensing \cite{lopaeva2013,Degen17} and quantum imaging.

In recent years, there has been a significant progress in the field of single-photon sources and their use in metrology, also thanks to the research activities started in National Metrological Institutions aiming at developing metrological techniques for the raising quantum technology applications exploiting single-photons.
In particular, single-photon sources were calibrated in a traceable way with respect to their absolute optical radiant flux and spectral power distribution
\cite{Rodiek17, Vaigu17}. Although this might be considered as a huge step towards the realization of a new photon standard source, the photon fluxes are still too low, and the purity is still insufficient for real practical use. 

Indeed single photon sources, typically, present intrinsic single-photon emission losses due to non-radiative decay mechanisms allowing the spontaneous decay of a single excited quantum emitter, without the emission of the expected single photon \cite{Migdall13,Eisaman11}. These non-radiative mechanisms dissipate the energy associated to the quantum emitter decay in some other way, for instance in the form of phonons in solid state system. In some cases, it is possible to effectively suppress such non-radiative decays (e.g. operating the emitter at cryogenic temperature), obtaining an almost ideal single-photon source showing nearly 100\% quantum efficiency \cite{Aharonovic2009, Müller2012, Brokmann2004}. Even in the presence of such ideal single photon sources, the photon flux emission is often limited by the poor optical collection and out-coupling of the emitted single-photons.

To enhance the photon fluxes of single photon sources, and make them more close to a deterministic behavior, photonic structures for enhancing the photon collection efficiencies are required.
%, because of natural isotropic emission of color center in (nano)diamond, quantum dots, molecules etc. .
Indeed, even exploiting high numerical aperture objectives, the collection efficiency is the order of the few percent. Several photonic structures for photon collection enhancement are under investigation, between them solid immersion lenses, waveguiding structures, micro-resonators or planar optical antennas \cite{Gerard98, Lee11, Claudon10, Reimer12, Munsch13, Gatto14, Ates12, Riedel14, Checcucci17, Loredo16, Somaschi16}.

The actual single-photon sources are still far from being completely predictable, deterministic and indistinguishable and more development is required to bring close-to-ideal single-photon sources, but they can be already exploited in several interesting applications in the field of quantum technologies in general, and for quantum imaging applications in particular, because of their intrinsic anti-bunching or sub-shot noise properties \cite{Chu17}.

On the contrary, non-radiative and optical losses as well decoherence limit the applicability of non-ideal sources to scenarios where deterministic emissions of ideal and completely identical single-photons are necessary; such as, e.g., when interference/interaction between two (or more) single-photons is exploited  as in quantum computation or simulation. Moreover, for the metrological point of view, ideal single photon sources on demand are highly desirable for developing absolute sources with finite photon number emission without intensity fluctuation. This could lead to a redefinition of radiometric and photometric unities in terms of number of photons (quantum Candela) \cite{Zwinkels2010}.

We highlight that a reasonable approximation of single photon sources are the heralded photons produced by SPDC \cite{Brida2011-0,Krapick2013}. As said in Sec. \ref{Twin-beam}, photons are always emitted in pairs with low probability, but one can get rid of the vacuum component since the detection of one photon of the pair heralds the presence of the other one, and the probability of a double pairs emission remains very low. These pseudo single-photon sources, based on PDC  or FWM, have interesting performances in terms of photon rate \cite{Polyakov09} and quality of the single photon emission, but they typically need temporal post-selection \cite{Polyakov09} to distinguish the heralded photon from a dominant unheralded background photon (even if there are remarkable exceptions \cite{Brida2011-0, Brida12c}).

Despite the fact that temporal post-selection is a limiting factor for the possibilities of practical exploitation of heralded single-photon sources in quantum technologies, this approach has been demonstrated recently for quantum enhanced absorption measurement and spectroscopy of a biological sample (hemoglobin)  with post-selection of the heralded single photons \cite{Whittaker2017}, and with selection performed by active feed-forward enabled by an optical shutter \cite{sabines2017}.

When it is possible to exploit temporal post-selection, e.g. for pulsed TWB emission, we should note that TWB, multi-modes or in the high gain regime, can be exploited as an heralded Fock state source, having on the heralding arm a photon number resolving detector with (nearly) ideal unit efficiency.

\section{Quantum imaging}\label{Quantum imaging}

Quantum imaging is one of the first application of quantum photonic technology \cite{Kolobov-2007}, where the quantum properties of light are exploited to enhance some peculiar aspect of the image formation process.

\subsection{Super-resolution with single photon emitters}\label{Super-resolution with single photon emitters}
Fluorophores and markers used in biological microscopy can be chosen among single photon emitters such as quantum dots, dye molecules or NV centers in (nano)diamond.  A single photon emitter  presents, by definition, a strong anti-correlation in the temporal photon emission, since at any time the presence of more than one photon is prevented by the excitation-decay process (see Fig. \ref{HBT_HOM_Fig} and its discussion in the text).

This effect is known as anti-bunching and can be exploited in many applications, from quantum information and communication to metrology and super-resolution imaging.
A possible approach exploits the measurement of the higher order Glauber correlation function at $\tau=0$,
\begin{equation}
g^{(k)}(\tau=0)=\frac{\langle\prod_{i=0}^{k-1}(\hat{n}-i)\rangle}{\langle \hat{n} \rangle^k},
\end{equation}
in each position of the image plane, obtaining an increase of the resolution of the single photon emitters map. Indeed if two or more single-photon emitters are closer than the diffraction limit (DL), thus indistinguishable from the standard fluorescence intensity map, the presence of coincident single-photon detection indicates the presence of more than one single-photon emitter.  This additional information, together with the intensity map, allows to reconstruct a super-resolved map \cite{Schwartz12,Schwartz13,Monticone14}. In general, for an arbitrary number of centers in the cluster, a scaling factor of $1/\sqrt{k}$ is achieved, when the Glauber autocorrelation function is measured up to the  $k$-th order ($g^{(k)}$).

\begin{figure}[tbp]
	\center
	\includegraphics[width=5in]{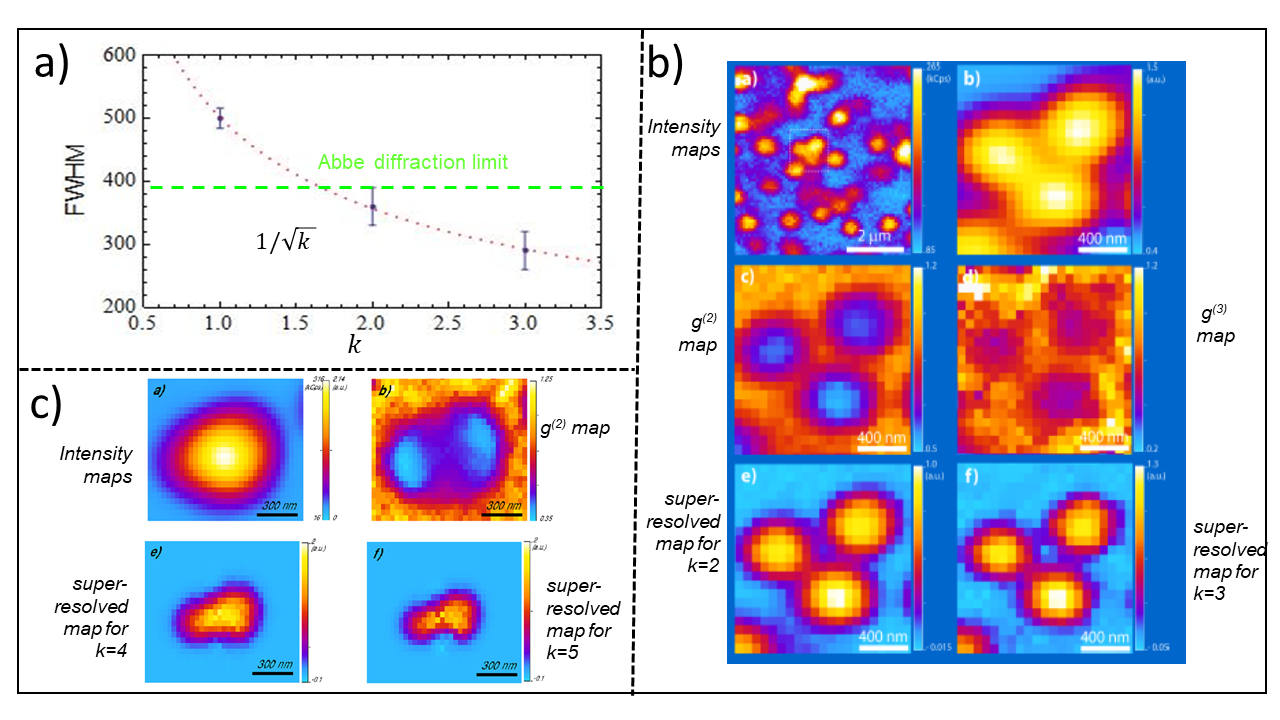}
	\caption{Super-resolution by exploiting anti-bunching in confocal single photon microscope \cite {Monticone14}. (a) experimental scaling of the FWHM of the point spread function with the order of the coincidence $k$. The green dashed line is the theoretical DL. (b) From the top to the bottom: Intensity map showing three NV centers at the limit of the DL separability, two- and three- fold coincidences maps, super-resolved map using the extra information gained by the measurement of $g^{(2)}$ and $g^{(3)}$. The white bar is 400 nm long. (c) Two emitters are at a distance below the DL and cannot be distinguished by looking at the intensity map. From the $g^{(2)}$ function (two-fold coincidence of photons) the presence of two dips suggests the existence of two distinct centers and this information is used in the super-resolved images (it is safely assumed that $g^{(k>2)}=0$). The black bar is 300 nm long.}\label{SubDiffraction_Fig}
\end{figure}

Experimental demonstrations of this technique has been performed also in wide-field and in confocal microscopy \cite{Schwartz13,Monticone14}. Fig.\ref{SubDiffraction_Fig} presents some of the results obtained in \cite {Monticone14}, by exploiting NV-centers in a single photon confocal microscope.

Recently it has been proposed to combine this high order correlation function method with the use of structured excitation light \cite{Classen17}. It has been shown that in this case one could reach a resolution increase of $k+ \sqrt{k}$ with respect to the DL.

Another technique, realized in 2017 \cite {Israel17}, exploits anti-bunching together with the natural "photo-blinking" effect found in many types of fluorophores and a fiber bundle collecting light with certain spatial resolution. After the reconstruction, a final resolution of 20 nm has been reported.

It is worth to mention the research on sub-diffraction imaging triggered by the interesting paper by Tsang and coworkers \cite{Tsang16}.  In their quantum-estimation-inspired proposal \cite{Helstrom76}, they observe that, given only two single photon emitters, a lot of information is present in the sum of the electromagnetic field of the two emitters at the image plane rather than just the sum of the intensity alone. Thanks to this, and using the quantum estimation theory, they were able to devise imaging techniques able to beat by far the diffraction limit. The idea of identifying the ultimate bound of the ability to estimate the distance between a pair of closely separated sources, achieving near-quantum-limited performance has inspired a relevant amount of theoretical researches (see e.g. \cite{Lupo16,Nair16,Rehacek17,Yang17}), and also a certain amount of proof-of-principle experiments (see e.g.  \cite{Tang16,Nair16b,Yang17b,Tham17,Paur16}). It is important to notice that all these results are limited to the case of only two sources; it is far from being obvious that similar performances can be achieved in the case of three or more sources. It is expected that this becomes an active research field in the future years also because of its inherent connection with multi-parameter quantum estimation problem \cite{Helstrom76,paris2009}.

\subsection{Sub-Shot-Noise Imaging}\label{SSNI_Sec}\label{Quantum enhanced displacement sensing}

\begin{figure}[thbp]
	\centering
	\includegraphics[ width=0.8\textwidth ]{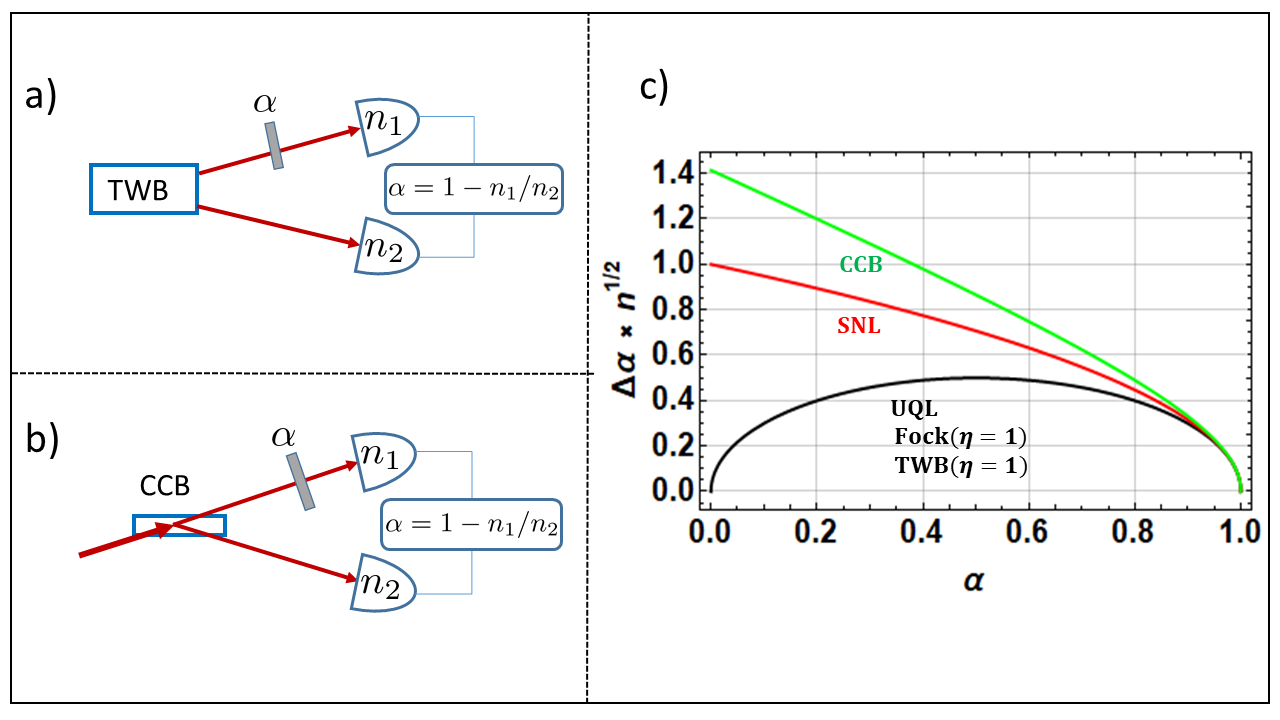}
	\caption{Loss estimation with correlated beams. (a) A pair of quantum correlated beams, i.e. twin-beam (TWB), are used: one beam is the probe interacting with the sample, while other beam is the reference. (b) Classically correlated beams (CCB), e.g. obtained by splitting a thermal beam, are exploited. (c) Uncertainty in function of the loss parameter $\alpha$ for TWB (ideal detection efficiency $\eta$) and CCB case compared to the shot noise limit (SNL) and to the ultimate quantum limit (UQL). }\label{D_alpha_TWB_Fig}  
\end{figure}

Extracting relevant information about a delicate sample with the lowest photon dose, is of paramount importance in biochemistry and biology to ensure that the processes being investigated are not shifted to an alternate pathway due to environmental stress.
Wide field microscopy is the simplest, fastest, less expensive and oldest imaging solution used, for example, for live-cell imaging and it has the advantage of requiring  the lowest photon dose, especially in transmission imaging with respect to scanning confocal microscope \cite{Cole:2015}.

The exploitation of multi-mode correlations in TWB has been proposed for high sensitivity wide field imaging of weak absorbing object in \cite{gatti2008} and a proof of principle of this technique has been reported by Brida \textit{et al.} in 2010 \cite{Brida2010}.
Recently, the first wide-field sub-SNL microscope \cite{Samantaray2017} has been realized, providing image of $10^{4}$ pixels  with a true (without post-selection) significant quantum enhancement, and a spatial resolution of few micrometers. This represents a considerable advancement towards a real application of quantum imaging. In the following we will discuss some details of this application.

We have already seen that sub-Poissonian light can be used to achieve SSN absorption estimation according to Eq. (\ref{uncdr}). However it is challenging to experimentally produce single modes with sub-Poissonian photon statistics. A different approach is to consider two correlated modes from a TWB state, and using one of them  as a reference for the noise in a ''differential'' scheme, as presented in Fig. \ref{D_alpha_TWB_Fig}(a).  In this case the uncertainty is:

\begin{equation} \label{Ualpha}
\Delta \alpha_{TWB}\simeq \sqrt{\frac{\alpha(1-\alpha)+ 2\sigma_{det}(1-\alpha)^{2}}{ \langle N_P \rangle}},
\end{equation}
where $\sigma_{det}$ is the the noise reduction parameter in Eq. (\ref{NRF_eta}). For TWB  it  equals the losses in detecting correlated photons, $\sigma_{det}=1-\eta$.  Ideal lossless detection leads to the ultimate quantum limit, $\Delta \alpha_{TWB} =  \sqrt{\alpha}\Delta \alpha_{SNL}\equiv\Delta \alpha_{UQL}$, which is represented by the black curve in Fig. \ref{D_alpha_Fig}(c).  The performance of TWB in loss estimation has been theoretically discussed in \cite{rarity,Hayat1999,Moreau2017,Losero2018} and demonstrated experimentally in \cite{Ta91, Moreau2017, Losero2018}. In \cite{Losero2018} more than 50\% sensitivity improvement with respect to the SNL has been reported, and it reached a 100\% improvements when compared with conventional two-beams approach, represented in Fig. \ref{D_alpha_TWB_Fig}(b). Note that the two-beams approach is extensively used in standard devices like spectrophotometers, where a classical beam is split in two by a beam splitter and one beam is used to monitor the instability of the source and detectors and to compensate for them. However, in classical correlated beams (CCB) generated in this way, only the super-Poissonian component of the fluctuations is correlated (sometimes called classical "excess noise"),  whereas the shot noise remains uncorrelated and cannot be compensated. Indeed TWB represent the natural extension of this method to the quantum regime. Recently similar schemes, but exploiting four wave mixing in Rubidium vapour, have been applied for monitoring index refraction change in plasmonic sensors \cite{lawrie2013, pooser2016}.

\begin{figure}[thbp]
	\centering
	\includegraphics[ width=0.8\textwidth ]{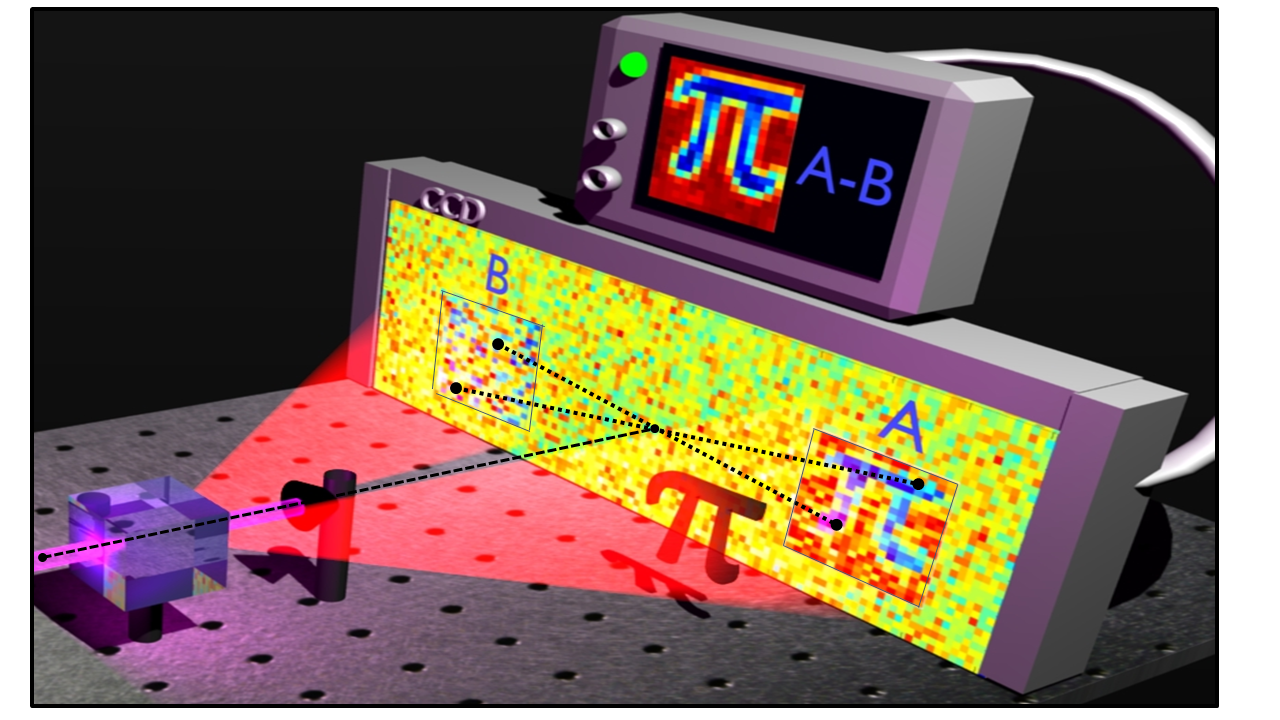}
	\caption{Simplified representation of a sub-shot-noise imaging scheme. The quantum noise in the probe beam, where a faint absorbing object is placed, can be removed by subtracting from the signal measured in each pixel of the image  the noise measured in the corresponding symmetric pixel in the correlated beam}\label{SSNI_Fig}	
\end{figure}
\begin{figure}[tbp]
	\centering
	\includegraphics[ width=0.8\textwidth ]{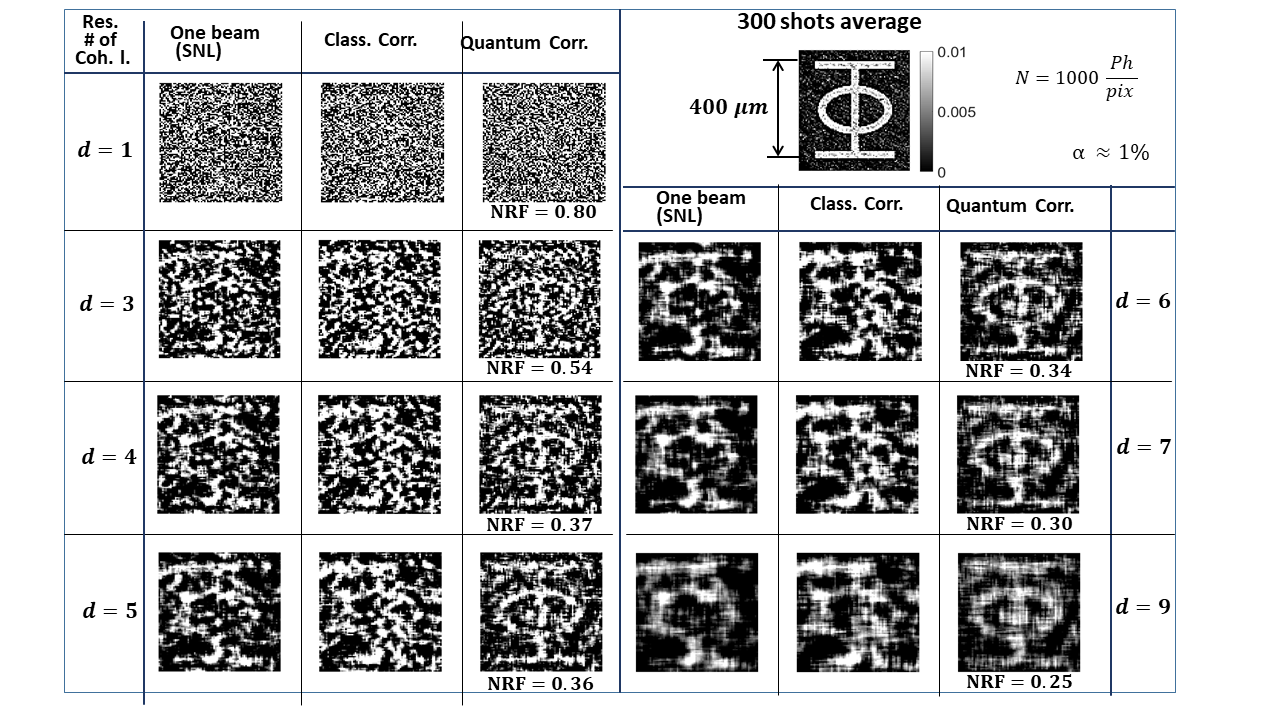}
	\caption{Experimental single-shot images extracted from  \cite{Samantaray2017}.  Direct classical imaging,  imaging exploiting classical correlation and imaging exploiting quantum correlation are compared. $d$ is the spatial resolution parameter, such that the effective resolution is $d*L_{coh}$, where $L_{coh}=5\mu$m is the spatial coherence length of the TWB correlation at the object plane.  The mean number of photons detected per pixel per frame is $N\sim1000$. The upper-right panel is the object image averaged over 300 shots. }\label{PhiSSNM_Fig}
\end{figure}

The extension of the technique from the two-mode case to the multi-mode case allows to realize wide field SSN imaging, as represented schematically in Fig. \ref{SSNI_Fig}. Basically, it is possible to match the pixel structure of a spatially resolving detector with the spatial distribution of the multi-mode TWB generated in the SPDC process, so that each pair of correlated spatial modes is precisely and entirely detected by a pair of pixels. The quantum correlation in transverse momentum, between $\textbf{q}$ and $\textbf{-q}$, is converted in spatial correlation among symmetric pixels in position $\textbf{x}$ and $\textbf{-x}$ at the object plane by a far field lens (not shown in the figure).
The noise of the image taken in one branch is removed pixel-by-pixel by "subtracting" the noise pattern measured on the other branch \cite{Brida2010}. In \cite{Samantaray2017}, the method has been demonstrated in a microscope configuration, using a  CCD camera operating in linear regime, with high quantum efficiency (95\%) and low (few electrons/ pixel/frame). The experimental evidence of the noise reduction and the sensitivity improvement obtained is presented in Fig. \ref{PhiSSNM_Fig}.

The spatial resolution of this technique is given by the traverse size of the correlation (coherence area), namely the uncertainty in the relative propagation direction of correlated photons, which is proportional to the inverse of the pump width. Pixel size should by large enough to include at least a coherence area, otherwise correlated photons may be lost affecting the noise reduction factor. In fact, Fig. \ref{PhiSSNM_Fig} shows that at full resolution ($d=1$) the noise reduction is modest while it becomes better if the resolution decreases.

Finally, we mention that differently proof of principles of quantum enhanced phase-contrast scanning microscopy exploiting  NOON states (N=2)\cite{On, Is, Wo, Cr} have been reported, although, in most of the cases, a significant enhancement without post-selection or losses compensation is still missing.

\subsection{Quantum enhanced displacement sensing}\label{Quantum enhanced displacement sensing}
\begin{figure}[tbp]
	\centering
	\includegraphics[ width=0.8\textwidth ]{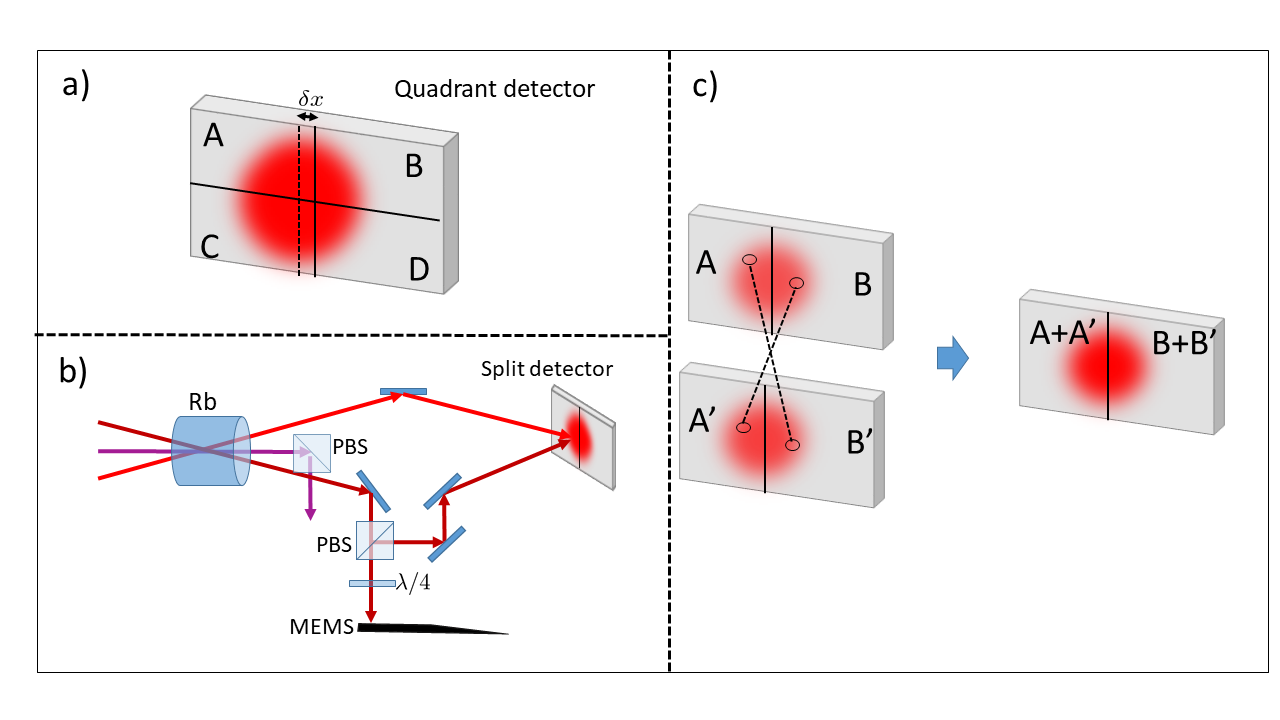}
	\caption{Beam displacement measurement. (a) Quadrant detector configuration. (b) Differential beam position measurement with quantum correlated
		TWB from four-wave mixing in Rubidium (Rb). The probe is sent
		to a polarizing beam splitter (PBS) where it passes through a quarter-wave
		plate before being focused onto a micro cantilever (MEMS)
		On the return path the probe beam is separated at the PBS and sent
		to a split detector toghether and overlapped to the reference one. (c) Scheme of the spatial correlations at the split detector}\label{Beam_Disp_Fig}
\end{figure}

Sub-shot-noise photon number/intensity correlations enable detecting displacement of a light beam with accuracy surpassing standard methods based on laser beams.
The standard way to monitor the wandering of a light beam is to send it to a quadrant detector as depicted in Fig. \ref{Beam_Disp_Fig} (a). This configuration is used  for measuring small displacements in many applications, for example, in atomic force microscopy, ultra weak absorption measurement, or single molecule tracking in biology \cite{Taylor-2014,Taylor-2013}. When the  beam is perfectly aligned to be symmetric with respect to the origin of the axis, the difference of the power measured in the different sectors is, on average, null. Small displacement in one direction, for instance along the $x$-axis will give a proportional change in the the photo-current difference \cite{Barnett2003},
\begin{equation} \label{beam_disp}
\langle n_{A+C}-n_{B+D}\rangle= \frac{2}{\pi^{1/2}} \frac{\delta x}{w}N,
\end{equation}

where $w$ is the waist of the beam (let us suppose a gaussian beam) and $N=n_{A+C}+n_{B+D}$ is the total photon number in the beam.   However,  due to the shot-noise which is independent between right and left sectors, the noise on the right  side of  Eq. (\ref{beam_disp}) is exactly $N^{1/2}$ and the minimum detectable displacement becomes:

\begin{equation} \label{D_x_SNL}
\Delta x= \left(\frac{\pi^{1/2} w}{2}\right) \frac{1}{\sqrt{N}}.
\end{equation}
It is interesting to note that the spatial resolution limit in this case is given by the energy available in the beam, or the power limit supported by the quadrant detector.

Multi-mode quantum correlations allow to beat the SNL in Eq.(\ref{D_x_SNL}). Following previous theoretical investigation \cite{Fabres2000}, the first demonstration of SSN displacement detection along one dimension has been obtained in \cite{Treps2002}, by combining vacuum squeezed beam and a coherent beam that are spatially orthogonal. Although the
resultant beam is not squeezed, it is shown to have strong internal spatial correlations between the portions of the beam impinging different sectors of the detector. The technique has been extended at the $2D$ case in \cite{Treps2003}.

An evolution of this technique has lead to the experimental demonstration of sub-shot-noise particle tracking in living system and  microrheology within Saccharomyces cerevisiae yeast cells \cite{Taylor-2013}. Lipid granules were tracked in real time as they diffuse through the cytoplasm  surpassing  the SNL of 42\%.
In typical laser-based particle tracking, the presence of a particle causes light to be scattered out of an incident field. The interference between scattered and transmitted fields leads to a  deflection of the incident field proportional to the displacement of the particle. The difficulties of using quantum correlation in this context are that, on one side such measurements are typically conducted at low frequencies where technical noise is dominant with respect to the shot noise, and on the other side, the distortion of the spatial mode propagating through high-numerical optical system and biological samples prevents the quantum light from matching the detection mode (quadrant detector).  To circumvent these issues, in \cite{Taylor-2013}  two separate fields are used, one for interrogation and the other, a ''flipped'' Gaussian local oscillator, to define the "detection mode". Interference among them, measured by a single photodiode, provides particle information equivalent to a quadrant photodiode.

%More in detail, a Gaussian probe field propagates transversely to the optical trapping axis, interrogating the particle and producing scattering, while a ‘flipped’ Gaussian local oscillator field, with a phase shift applied to one half of its transverse profile, propagates along the trap axis and acts to define the detection mode (see Fig. 2). Direct detection of the interference between the flipped local oscillator and scattered light on a single photodiode provides particle position information equivalent to the quadrant photodiode in standard particle tracking.

In 2015, Pooser and al. demonstrated a noise reduction of 60\% below the SNL in state-of-the-art displacement sensitivity for a MEMS cantilever \cite{Pooser2015}. They used a quantum beam of power up to hundreds of $\mu$W reaching   few fm$/\sqrt{\textrm{Hz}}$ resolution in the region of $10^5$ kHz. Fig. \ref{Beam_Disp_Fig}(b) presents the sketch of the experiment where two spatially multi-mode correlated TWB, generated by FWM, are superimposed at a split detector. As we have discussed in Sec. \ref{SSNI_Sec},  the variance of the photo-current difference between symmetric portions of the two beams in Fig. \ref{Beam_Disp_Fig}(c), either A and B' or A' and B, is lower than the SNL. Moreover, this is true also when they are spatially overlapped at the same detector. Therefore the fluctuations of left hand side of the Eq. (\ref{beam_disp}) are reduced below the SNL and this is the origin of the advantage of the technique.

We mention also that a similar scheme has been investigated in the discrete case, considering spatially entangled biphoton pairs \cite{Lyons2016}. It has been found that in principle the smallest resolvable parameter of a simple split detector scales as the inverse of the number of pairs (Heisenberg scaling) when the last is very small.

In a different perspective, but conceptually analogue to the previous case, there is the  estimation of the displacement of a shadow casts by a fully opaque object intercepting a probe beam. In \cite{Toninelli2017}, a noise reduction corresponding to an improvement in position sensitivity of up to 17\% has been obtained with bi-photon pairs, created with SPDC and detected by an EMCCD camera employed as a photon-number-resolving split-detector. As reported in literature \cite{Avella2016}, EMCCD can be  operated as a photon number resolving detector, thanks to spatial multiplexing, where each pixel is operated in "on-off" regime by setting a discriminating threshold.

\subsection{Quantum Ghost imaging and spectroscopy}\label{Quantum Ghost imaging and spectroscopy}

Quantum Ghost imaging (QGI) and Ghost-spectroscopy (QGS) are two complementary techniques based on signal and idler photons correlations in SPDC (or FWM).

Quantum Ghost imaging has been proposed in 1994 \cite{belinskii} and experimentally realized in 1995 by Pittman et al \cite{pittman}. It makes possible to reconstruct the transmittance(or reflectance) profile of an object, placed in the signal beam, although the light after the interaction is collected by a single pixel or  "bucket" detector, i.e. without spatial resolution. The spatial information is retrieved by performing correlation measurements (e.g. time coincidence) among the single pixel detector and each pixel (or position) of a spatially resolving  (or scanning) detector on the idler beam. The working principle is the following: the spatial selection performed in the idler beam automatically identifies a small range of allowed position for the correlated signal photons at the object plane, expressed by a function $\Gamma(x_{i}+ x_{s})$ peaked into zero and  which has to be narrower than the minimum size of the object variation scale. If a coincidence is registered among the two detectors, this means that in that specific point the object is transmitting. On the other side, if the coincidence is missed, the object has absorbed the signal photon. In this way, by measuring the number of coincidences $R(x_{i})$  in function of the position $x_{i}$ of the idler pixel, it is possible to reconstruct the whole transmission $T_{s}$ of the object in the signal path,   $R(x_{i})\propto\int_{Bucket} \Gamma(x_{i}+ x_{s}) T_{s}(x_{s}) dx_{s} \simeq T_{s}(-x_{i})$. Instead of measuring time coincidences between SPADs operated in on-off modality, it is also possible to measure temporal correlation of the intensity fluctuations among a couple of linear detector if they are sensitive enough to register such signals, which are usually faint.

\begin{figure}[htbp]
	\centering
	\includegraphics[ width=0.8\textwidth ]{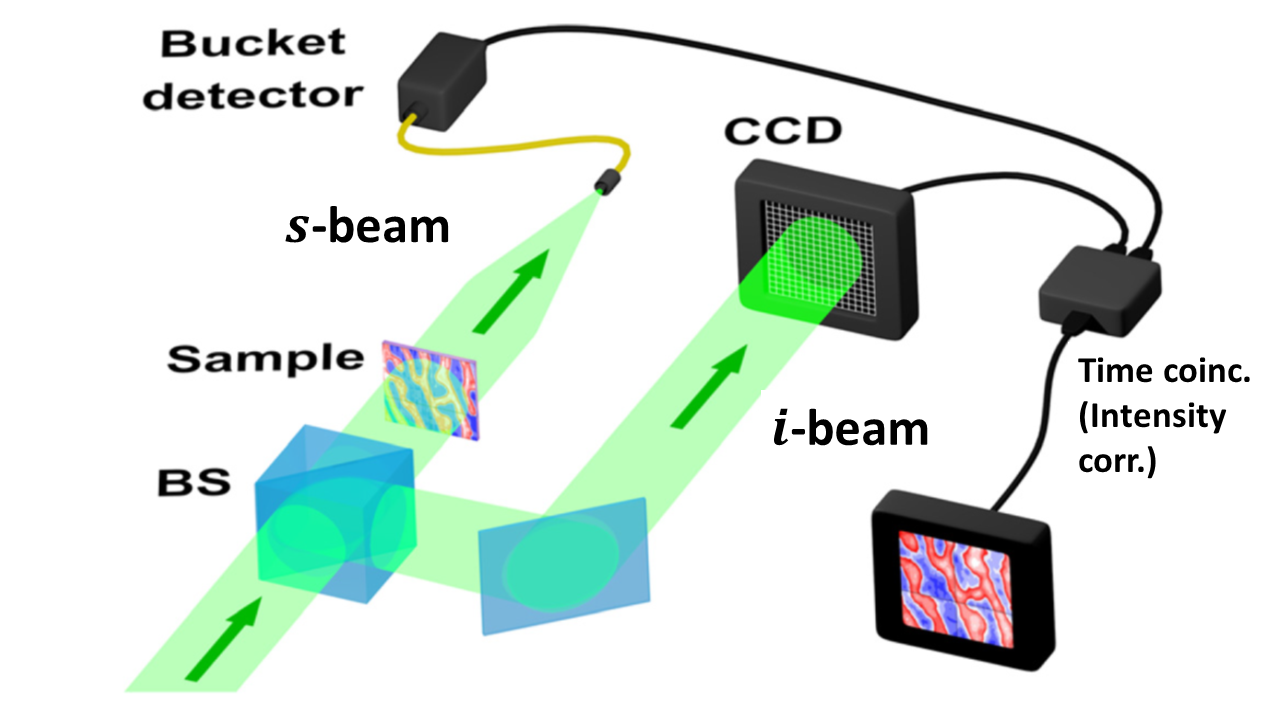}
	\caption{Sketch of a typical thermal ghost imaging set-up: A single beam with thermal fluctuation in space and time is divided by a Beam splitter (BS). The signal and idler branches are imperfect copy of the same beam and therefore they share some degree of correlation.  Due to the brightness of the thermal sources, correlation among the intensity fluctuation of the photo-currents are usually evaluated for the reconstruction, instead of the coincidences. [See the original picture is \cite{Meda2015} ] }\label{QuGI_ThGI_Fig}
\end{figure}

A lively debate, whether ghost imaging truly requires quantum light has started just after its first demonstration. Many theoretical investigations \cite{gatti,bache,Shapiro} and experiments \cite{bennink,ferri,valencia,chen,zhai} have demonstrated that almost all the quantum ghost imaging features can be mimicked by classical correlation, usually generated by splitting an spatially incoherent single pseudo-thermal beam (see Fig. \ref{QuGI_ThGI_Fig}) obtained from a laser beam scattered by a rotating ground-glass disk. However, thermal GI can be obtained with sunlight \cite{liu}) or broadband super-luminescent diode source \cite{Hartmann2015} although the short coherence time make the detection scheme rather difficult. An interesting compromise between thermal-light-based and TWB-based GI can be found in \cite{Puddu2007}, where the authors used speckles correlation in intense TWB seeded with thermal light.

Even computational technique allows GI reconstruction \cite{shapiro1,bromberg,katz}, by impressing random spatial patterns on a single beam modulated in a controllable and known way (spatial light modulators, either in phase or amplitude are standard devices currently used in many imaging applications).

Since the first proof-of-principle demonstrations, either with TWBs or  with classical light, substantial advancements towards the applicability of GI have been achieved: drastically improving the SNR by differential GI \cite{bina}, measuring reflected photons \cite{meyers2008}, using polychromatic light \cite{welsh2013,duan2013}, and extending GI to 3-dimensional reconstruction \cite{sun2013}. For enhancing visibility and contrast of GI, the use of high-order correlations has been proposed and applied \cite{chen,Chan2010,Allevi2012}. Also compressive sensing technique has been successfully transferred to GI, speeding up the image reconstruction \cite{katz,Yu2014}.

There is a really huge amount of literature on GI, both concerning its relation with more fundamental questions about the quantum-classical boundary and  applications. We do not intend here to be exhaustive and we advise the interested reader referring to dedicated reviews \cite{Shapiro, moreau2018}.

Among the general advantages of ghost imaging which have intrigued scientists there is robustness against effects like scattering and phase distortion which are experienced between the object and the bucket detector. This can have important applications in an open-air ranging system or for inspecting biological
samples, where tissues represent the diffusive medium. In fact, even if the propagation directions of photons are scrambled after the sample, it does not make any difference for the bucket detector which integrate in any case. Many works have demonstrated turbulence free ghost imaging and ghost imaging through turbid media \cite{bina, dixon2011,gong2011,meyers2012}.

 Another possibility offered by quantum ghost imaging that is difficult to emulate with classical sources is the "two-color" GI \cite{rubin,chan2009PRA,karmakar}. By producing pairs of non-degenerate photons, the object can be illuminated by photons with a significantly different wavelength from the ones detected with spatial resolution on the second beam. For example,  image reconstruction in the infra-red range can be achieved with cameras operated in the visible spectrum \cite{aspden2015}.
 Moreover, GI is always an option in all the situations in which, space constraints or  "hostile" environment conditions, such as e.g. extreme temperature or high electromagnetic fields not allowing the set-up an imaging system with a spatially resolving detector close to the sample. As proof of principle, in \cite{Meda2015} the authors apply the GI approach to make the image of the magnetic Faraday effect in the domain structure of a  yttrium iron garnet sample, opening possible development of magneto-optical-imaging at extreme cryogenic temperatures  or in presence of high magnetic fields.

In conclusion, one of the big advantage in using quantum correlation is given by the ability of rejecting external technical noise and, in general, uncorrelated background, especially when faint light levels are required \cite{morris2015,Erkmen2009,brida2011-2}. Furthermore, note that the noise can be also due to uncorrelated photons produced by the same source used for GI. Indeed, all the spatio-temporal modes collected by the bucket detector which are not correlated with the spatial-mode impinging on a specific pixel of the spatially resolving detector contributes to the uncorrelated background.

Basically, since the photons are produced in pairs at the same time the coincidences among two SPADs appear within a very short time window $\Delta t$ (typically few $ns$, because of the detectors and correlation/coincidence-electronic jitter). The number of "true" detection events seen by each detector is simply given by $N^{(T)}_{i(s)}=\eta_{i(s)}P \Delta t$, where $P$ is the pairs production rate. Analogously, the background counts are  $N^{(B)}_{i(s)}=B_{i(s)}\Delta t $ with $B_{i(s)}$ is the rate of background counts.
The  number of true coincidences is proportional to the probability of joint detection by the two single-photon detectors (located in the two channels) $R^{(T)}=\eta_{i}\eta_{s} P \Delta t$.  Assuming that noise photocounts of the two single-photon detectors are uncorrelated, the number of accidental coincidences $R^{(A)}$ is given the product of the probability components due to accidental overlap of SPDC and noise photocounts, $R^{(A)}=N^{(B)}_{i} N^{(B)}_{s}+N^{(B)}_{i} N^{(T)}_{s}+N^{(T)}_{i}N^{(B)}_{s}=[B_{i}B_{s}+(\eta_{i}B_{i}+\eta_{s}B_{s})P](\Delta t)^{2} $, where we observe that it scales as the square of the detection time window. Let us suppose that the background rate is dominant, i.e. $N^{(B)}_{i(s)}>> N^{(T)}_{i(s)}$.
The signal to noise ratio of quantities proportional to the true coincidence is  $SNR_{QGI}=R^{(T)}/R^{(A)}\simeq\eta_{i}\eta_{s}P/(B_{i}B_{s}\Delta t)$. This performance should be
compared with the direct single branch case in which the SNR is $SNR_{D}=N^{(T)}_{s}/N^{(B)}_{s}\simeq\eta_{s}P/B_{s}$. Therefore, the advantage in terms of SNR of the quantum correlation technique is $SNR_{Q}/SNR_{D}=\eta_{i}/(B_{s}\Delta t)$ \cite{Kalashnikov2014}.
\textit{This means that the QGI delivers better image quality reconstruction when the efficiency of the reference (idler) arm is high and the number of photon in the detection time is smaller unity}. This remains valid also in the case of analog detection, in which rather than temporal coincidence, intensity correlations are measured, as derived theoretically and demonstrated experimentally in \cite{brida2011-2}.
In \cite{morris2015}, QGI of a wasp wing has been obtained by less than 0.5 photon detected per pixel, with the help of a dedicated compressive sensing algorithm. In that case an intensified CCD camera was gated by the photons arriving at a SPAD, which  plays the role of the single pixel detector. An imaging preserving delay line is necessary to compensate for the time necessary to the electronics for sending the triggering input to the camera.

Similarly to ghost imaging, in quantum QGS, the correlation in frequency shown in Eq. (\ref{phasematch}) allows to retrieve the absorption spectrum of $T_{s}(\omega_{s})$ by measuring the temporal coincidences among photon pairs $R(\omega_{i})\propto T_{s}(\omega_{p}-\omega_{i})$, and the spectral selection can be done in the beam which does not interact with the object. The first proposals and experiments have been reported in \cite{Scarcelli2003} and \cite{Yabushita2004}. This scheme provides the same advantages of ghost imaging, offering the possibility of spectral selection in range significantly different  with respect to the one of the light impinging the sample \cite{Slattery2013}. The second benefit is that the spectral profile can be obtained by exploiting an extremely low signal level compared to external and technical noise \cite{Kalashnikov2014}. Again, it  is very promising in all those applications in which delicate or photosensitive systems require the lowest dose of photons, such as in biological spectroscopy.
We mention that, even ghost spectroscopy can be realized with classical light, but in a completely different regime \cite{Janassek2018}.
\subsection{Other advanced quantum imaging protocols}

\subsubsection{Imaging and spectroscopy without photons detection:}

We have seen that QGI and QGS allow one to measure the spatial and spectral properties of a system without spatial and spectral selection of the photons interacting with it. More complicated schemes, which use the non-linear interference occurring in non-linear crystal, allow even to eliminate the necessity of detecting those photons at all \cite{Kalashnikov2016}. A non-linear interferometer is equivalent to a Mach-Zehnder one, where the two beam splitters are replaced by two non-linear crystals  \cite{Chekhova2016} . The signal and idler beams are produced in the first crystal from the pump beam. The peculiar  feature of non-linear interference is that it involves a precise phase relation between all three propagating photons, i.e. the signal, the idler and the pump. Only if they have the right phase relation an output from the second crystal is observed. This differs from conventional interferometry, where the interference pattern is defined by the phase of the signal photon. Measuring the output intensity of the idler beam after the second crystal allows one to infer the phase eventually acquired by the signal beam passing through a sample. Here, the detection of signal photons is not necessary. Similar process can open entire new metrology schemes in optical imaging and sensing. Non linear interference has been used in imaging protocols with undetected photons \cite{Lemos2014}, interferometry below the shot-noise \cite{Hudelist2014,Manceau2017}, hybrid atom-light interferometers\cite{Chen2015}.

\subsubsection{Quantum Illumination}

Quantum illumination (QI) is a protocol proposed in 2008 by Lloyd \cite{Lloyd-2008}  and soon reformulated exploiting TWB by Shapiro \cite{Tan-2008, Shapiro-2009}, providing a quantum improvement in target detection (radar like configuration) in  presence of a dominant thermal background. A probe beam is addressed to a region of the space where a weak reflecting target may be present or not. However, the partial reflection of the probe beam is hidden in a much stronger background. A quantum receiver performing a joint measurement between the reflected probe and an ancillary quantum correlated beam, allows to discriminate the faint signal component from the noise, revealing the presence of the target. Indeed quantum illumination with TWB delivers a 6 $\textrm{dB}$ (a factor 4) improvement in the error probability exponent with respect to the best classical strategy.

The outstanding feature of quantum illumination, which makes it unique in the panorama of quantum enhanced measurements,  is that its advantage does not depend neither on losses nor on the noise the probe experiences during the propagation and the interaction with the target. It is important to note that both these processes cause decoherence and therefore the initial entanglement or quantum correlation, is completely lost at the detection stage. This property is very valuable, especially in view of real world applications, where noise and losses are often unavoidable.

A first experimental realization of a quantum illumination-like protocol has been reported in \cite{lopaeva2013, Lopaeva-2014} considering a restricted scenario in which only  intensity measurements (phase-insensitive) were exploited. Here a photon number measurement is performed independently in the reference arm and in the probe arm, then the covariance of the two quantities is evaluated.  In this case, because of the high correlation in TWB,  unreachable by classical beams, the quantum advantage scales as $1+M/n=1+ 1/\mu$, where $M$ is the number of spatio-temporal modes (the inverse of the bandwidth) and $\mu$ is the mean number of photons in each mode. Interestingly, this corresponds to the increasing in the total mutual information between two parties sharing a quantum correlated states with respect to the one provided by classical correlations \cite{Ragy-2014}. In \cite{lopaeva2013, Lopaeva-2014} a 10 dB improvement in terms of SNR with respect to correlated thermal beams, has been achieved.

Quantum receiver able to beat the performance of the optimal classical strategy by 3 $\textrm{dB}$ have been realized in \cite{Zhang-2015}, based on optical parametric amplifier scheme \cite{Tan-2008, Guha-2009}. The advantage, however, requires that the returning probe, when present, has known amplitude and phase, which is not the case in many light detection and ranging (lidar) applications. At lidar wavelengths, most target surfaces are sufficiently optically rough that their returns are randomly distributed in amplitude and phase (speckles). In \cite{Zhuang2017PRA} the authors show that second harmonic generation process allows to enhanced detection of Rayleigh-fading targets, although with reduced (subexponential) advantage. The same receiver, in case of nonfading target, achieves QI's full 6 dB advantage \cite{Zhuang2017PRL}.  Further improvements can be, in principle, obtained by using photon-subtracted two-mode-squeezed states \cite{Zhang-2014}, although their practical realisation is extremely challenging.

Recently microwave/optical QI  has been proposed in \cite{Barzanjea-2015}. It would be of utmost importance to move from the optical demonstration to the microwave region because it is the natural domain in which QI could have vast application in the field of remote sensing.
Quantum illumination has potential application also in the field of secure quantum communication, for defeating passive eavesdropping attacks \cite{Shapiro-2009b, Zhang-2013}. The idea is that only authorized parties which share the original quantum correlations can achieve information on the modulation of an weak reflection, if enough noise is artificially added in the channel.

Quantum illumination can find also application in biological measurements when a delicate probing is essential and, at the same time, surrounding environment background can be of disturbance.

\section{Quantum Photometry}\label{Quantum Photometry}

One of the strongest motivation for developing single- and few-photon metrology is related to the investigation of the response of biological systems at few photons level.
Some biophysical and biochemical processes, for example phototransduction in vision, or photosynthesis,  are triggered by the absorption of a single or few photons.
The physical stimulus of vision is a consequence of the interaction between one photon and a family of light-sensitive proteins called opsins, typically found in the vertebrates photoreceptor cells present in the eye. Our visual system provides hundreds of millions of photoreceptors, divided in two families of photosensitive cells, rods and cones, that allow eye sensitivity to a huge range of luminance values (ranging from 105 cd/$\textrm{m}^{2}$ down to the single-photon level). This system converts the light pattern in an electrical pattern, later processed and sent to the optical nerve by other cells present in the retina (ganglion and interneurons). Under high illumination level, rods saturate and only cones, divided in three classes according to different spectral sensitivity, contribute to the vision (photopic vision). As the level of light decreases, rods start to be active (mesopic vision) together with cones, below a certain luminance level ($~10^{-2}$ cd/$\textrm{m}^{2}$), only rods are operative (scotopic vision) \cite{Zwinkels2010}.

Rods are recognized to act as photocounters with very high quantum efficiency and low dark noise \cite{Rieke1998}: evidence of the detection of few photons is present in literature since 1942 \cite{Hecht1942}. In this pioneering work, a Poissonian source of photons was considered and compared to the statistics of the response of a dark-adapted human observer, demonstrating that such an observer is able to detect very few photons (5-7 photons). However, the response of the retina at single or few-photon level is not easy to be investigated with classical light sources, dominated by shot-noise for low photon fluxes causing random multiphoton events. As discussed in this review, only quantum optical sources of light allow to overcome this limitation. The development of deterministic single-photon sources, based on time correlated photon pairs has recently enabled the investigation of the fundamental limit of the rods sensitivity \cite{Phan2014}, addressing the question if even a single-photon can be discriminated by the human being \cite{Tinsley2016}.

In particular, in \cite{Phan2014} a single rod toad cell held in a suction pipette is stimulated by photons produced by a SPDC heralded single photon source. Detection of a photon in the idler path by a SPAD drives an optical shutter in the signal path which allows only announced photon to address the photoreceptor. The registered amplitude of the electrical signal from the cell presents unambiguous evidence of the detection of some of the announced photons. The Authors also evaluated the quantum efficiency of the rod by the Klyshko two-photon technique to be $\eta_{rod}=0.29(0.04)$.

The first behavioural measurements in which single photons are sent to the human eye have been reported in \cite{Tinsley2016}. In that case a sophisticated psychophysical test is used,  called two-alternative forced-choice (2AFC) protocol. In summary, in each trial the subject has to identify a light stimulus (single photon) that can be delivered in two  separated temporal bins. The number of correct answers on which time bin contained the photon allows to exclude psychophysical false detection.
In that case the suppression of the multi-photon component has been obtained by post-selection of the trials in  with only one photon was detected by a EMCCD used as a spatially multiplexed photon number resolving detector in the idler path. The results show that humans can detect a single-photon incident on the cornea with a probability
significantly above chance. Moreover,  the probability of detecting a single photon is modulated by the presence of an earlier photon, suggesting a priming process that temporarily enhances the effective gain of the visual system on the timescale of seconds.

\begin{figure}[htbp]
	\centering
	\includegraphics[ width=0.8\textwidth ]{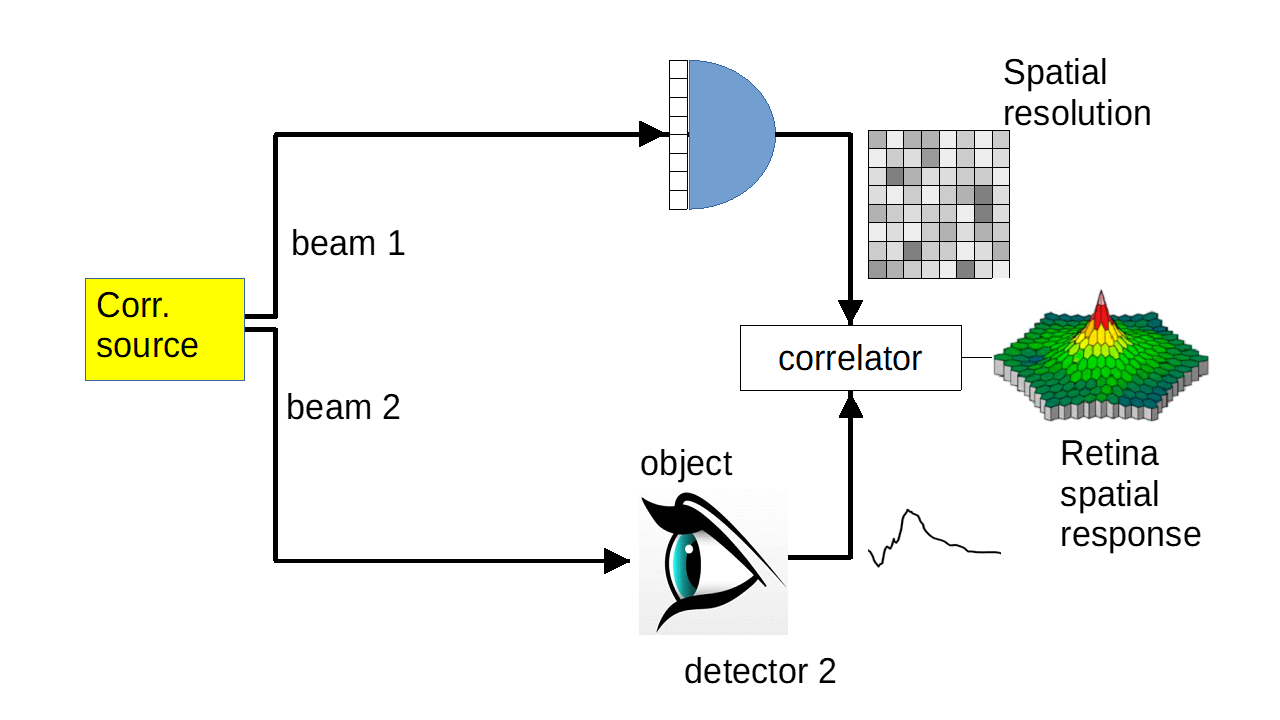}
	\caption{Possible quantum ghost imaging scheme applied to retina investigation. The retina is here the bucket detector providing an integrated signal. Using the ghost imaging approach it is possible to reconstruct the spatial mapping of its responce.}\label{RetinaGI_Fig}
\end{figure}

Even if single photon sensitivity of the human eye has been demonstrated, the mechanism of vision at very low-light condition and the transition to the mesopic regime remains mostly unexplored. This opens the opportunity to develop a broad new metrological field, we will refer to as quantum photometry. Indeed, applying new quantum radiometry tools can have a strong impact in understanding the physiological processes involved, eventually supporting vision neurophysiology and research in computer vision as well as the investigation and early detection of pathological conditions.  In addition, it must be stressed that  the human retina is part of the central nervous system (CNS) which can be directly in-vivo approached and that many CNS-related disorders (from multiple sclerosis to neurodegenerative disorders) are known to affect the retina as well, making the retina a powerful biomarker both of disease progression and of pharmacological treatments.
Daring a litte more, QGI and QGS have the potential to be applied to the retina (a sketch is presented in Fig. \ref{RetinaGI_Fig} ).  Retina spatial and spectral response could be reconstructed by the global evoked electric signal collected with  noninvasive skin electrodes, probably the less-invasive approach possible to the problem of retina spatial mapping and damaged area detection. Furthermore, QGI is expected to be particularly efficient and accurate with respect to standard techniques when faint illumination is used, down to the regime in which few- or even single- photons are detected one by one.

\section{Final Remarks and Conclusion}\label{Final Remarks and Conclusion}

Quantum photonics is considered one of the major future challenge, but also a big opportunity, for the forthcoming quantum industry with respect to innovation and high technology. Therefore, in the next future it is expected that national metrology institutes will be asked, by industries, standardisation bodies and governments, to contribute to standardization and certification of quantum photonic technologies \cite{Langer09, Alleaume14}. The metrological community should be proactive in promoting take-up of metrology in the development of these technologies \cite{Rastello14}. Photonics covers transversally a relevant part of the whole field of quantum technologies \cite{Acin18}: 	
\begin{itemize}		
		\item Quantum Sensing \& Metrology (with optical, magneto-optical and optomechanical imaging and sensing techniques, including new metrological standards),	
		\item Quantum Communication (with quantum key distribution, with discrete and  with continuous variables),		
		\item Quantum Simulation (with photonic circuits).		
\end{itemize}	
In particular, in this review we have presented and discussed Quantum Sensing \& Metrology techniques exploiting sub-Poissonian photon statistics and non-classical photon number correlations. We have highlighted state of the art non-classical light sources, such as single photon sources and twin beams, discussing their performance and limitations. We have seen that the non-classical states of light produced by these sources can represent a breakthrough, improving both the spatial resolution and sensitivity of measurements, especially in dim optical power condition, even though actual techniques are in several cases strongly limited by optical losses. This possibility of reducing the measurement uncertainty opens itself new research directions in modern optical metrology. Moreover, the development of quantum enhanced optical measurement is also a great opportunity for metrologists related to the above mentioned demand of characterization and certification infrastructure for quantum technologies. It is of particular interest, from the radiometric point of view,  the development of absolute light sources with sub-shot-noise performance in the few- and single-photon regime as well as reliable absolute calibration techniques for detectors based on quantum correlation. These future developments of deterministic quantum sources could be disruptive for radiometry and photometry, leading to the realisation of a new type of primary standard and paving the way for a possible redefinition of the unit Candela in terms of number of photons.  \cite{Zwinkels2010}. At the same time, the possibility of investigating biophysical process with these new tools, absolute and accurate, may start new metrological fields and applications, one example that we have reported here is the study of vision mechanisms at the single photon level, dubbed quantum photometry.

\ack
This work has been supported by EMPIR 14IND05 ``MIQC2'', EMPIR 17FUN01 ``BeCOMe'', EMPIR 17FUN05 ``SIQUST''   (the EMPIR initiative is co-funded by the EU H2020 and the EMPIR Participating States). IRB and IPD are deeply indebted with Elena Losero for useful suggestions to improve the manuscript.

\end{document}